\documentclass[11pt, a4paper]{article}
\usepackage[a4paper]{geometry} 
\usepackage{amsmath}
\usepackage{algorithm}
\usepackage{algpseudocode}
\usepackage{mathtools}
\usepackage{amssymb}
\usepackage{bm}
\usepackage{physics}
\usepackage{booktabs}
\usepackage{tabularx}
\usepackage{enumitem}
\usepackage{pifont}
\usepackage{subcaption}
\usepackage{float}
\usepackage{graphicx} 
\usepackage[affil-it]{authblk} 
\newcommand{\pderiv}[3]{\left(\frac{\partial #1}{\partial #2}\right)_{#3}}

\newcommand{\sderiv}[2]{\frac{\mathrm{D} #1}{\mathrm{D} #2}}

\title{How to Recover Oscillation-Free Pressure in Real Fluids: The RFQC Method and Its Liquid-Upwind Anomaly}
\author[1,2]{Haotong Bai}
\author[1,2]{Yixin Yang\thanks{Corresponding author: yangyixin@nudt.edu.cn}}
\author[2]{Wenjia Xie\thanks{Corresponding author: xiewenjia@nudt.edu.cn}}
\author[3]{Ping Yi}
\author[1,2]{Mingbo Sun\thanks{Corresponding author: sunmingbo@nudt.edu.cn}}

\affil[1]{Advanced Propulsion Technology Laboratory, National University of Defense Technology, Kaifu District, Changsha, Hunan 41073, China}
\affil[2]{College of Aerospace Science and Engineering, National University of Defense Technology, Kaifu District, Changsha, Hunan 41073, China}
\affil[3]{Institute of Power Plants and Automation, Shanghai Jiao Tong University, Shanghai, 200240, China}

\date{}
\begin{document}

\maketitle
\begin{abstract}
\noindent \textbf{Abstract~~~}
From the perspective of continuum thermodynamics, we revisit the pressure oscillation problem in finite-volume methods for multiphase real fluids and clarify the physical counterpart of the Real Fluid Quasi-Conservative (RFQC) method ~\cite{Bai2026RFQC}. The pressure oscillation in conservative finite-volume methods originates from their implicit thermodynamic equilibrium assumption, whereas recovering an oscillation-free pressure requires additional physical information. The RFQC method achieves this by evolving the affine parameters $\xi$ and $E_0$ of the isentropic internal-energy--pressure relation along pathlines, while the thermodynamic re-projection converts the deviation from the isentropic trajectory into an internal-energy error, thereby ensuring the thermodynamic consistency and numerical stability of the method. We then investigate the applicability limit of the RFQC method and identify a Liquid-upwind Anomaly (LUA) in extreme phase-change cases. For a Riemann problem involving liquid–vapor phase change, a numerical anomaly may occur if a liquid-upwind translational velocity is initially superimposed. Theoretical analysis reveals that this anomaly is initiated by the jump in the affine slope $\xi$ during phase change, which delays pressure rise in the downstream vapor cell. Concurrently, the re-projection removes the positive pressure increment, repeatedly generating large internal-energy errors and trapping the vapor cell in a cycle of delayed pressure recovery. The analysis indicates that the LUA is a start-up anomaly, which can be resolved by introducing a regularization strategy at the initial discontinuity. With the proposed regularization strategy, the RFQC method is equipped with enhanced accuracy and robustness for extreme thermodynamic flows, such as sonic phase-change jets.
\\
\textbf{Keywords~~~}Phase transition, Real fluid, Pressure oscillation, Finite volume method
\end{abstract}

\section{Introduction}

The pressure oscillation is a classical problem in computational fluid dynamics (CFD) of two-phase flows, multicomponent flows, and real fluids. Researchers have studied it extensively for many years and have developed a series of numerical methods and physical insights. In classical two-phase and multicomponent flows, pressure oscillations appear at vapor--liquid interfaces or contact interfaces. Researchers subsequently recognized that this problem arises because two-phase fluids cannot be simply averaged within a single finite-volume cell, with the pressure recovered via a single equation of state (EoS) \cite{Abgrall1996,Karni1994,Abgrall2001}. Based on this understanding, various numerical methods have been developed for two-phase and multicomponent flows. The first type assumes that pressure equilibrium in a finite-volume cell should involve physical relaxation. Additional equations are introduced to describe the vapor and liquid phases separately, and relaxation terms are used to drive the two fluids toward mechanical equilibrium at the interface. Typical examples include the well-known Baer--Nunziato seven-equation model \cite{BaerNunziato1986}, the Saurel--Abgrall seven-equation model \cite{Saurel1999}, and six-equation methods with pressure non-equilibrium relaxation \cite{Saurel2009,Pelanti2014}. The second type assumes that the fluid in a finite-volume cell has instantaneously reached pressure equilibrium, but this equilibrium is maintained by additional numerical treatments. These treatments can be further divided into two classes. The first class maintains pressure equilibrium through additional evolution equations, such as the volume-fraction equation in the Kapila five-equation model \cite{Kapila2001,Murrone2005}, and the additional transport equations proposed for the Euler equations with stiffened-gas EoS \cite{Shyue1998,Johnsen2006}, Mie--Gruneisen EoS \cite{Shyue2001}, and van der Waals EoS \cite{Shyue1999,Pantano2017}. The second class maintains pressure equilibrium by modifying the energy equation or the energy flux, such as the double-flux method \cite{Abgrall2001,Billet2003} and the pressure-based method \cite{Karni1994,Karni1992,Karni1996}.

In real fluids, although the pressure oscillation problem is essentially similar to that in classical two-phase flows, in the sense that fluids with different thermodynamic states cannot be directly and simply averaged, the problem becomes more challenging. Real fluids are described by complex EoS and their thermodynamic states may change sharply during transcritical and phase-change processes. Under the homogeneous equilibrium model (HEM) framework \cite{Brennen2005}, the discretized real fluid between any two adjacent finite-volume cells or time intervals can be viewed as a "two-phase flow" with completely distinct thermodynamic properties. This leads to two main difficulties in pressure recovery for real fluids. First, the phase difference cannot be clearly identified. At supercritical pressures, a fluid may directly change from a liquid-like state to a gas-like state without phase separation. During this transcritical transition, the traditional concepts of "vapor" and "liquid" no longer exist, which makes it difficult to naturally extend physical-relaxation methods for two phases to real fluids. The second difficulty is that complex equations of state for real fluids, such as the Peng--Robinson EoS \cite{Peng1976} and the Redlich--Kwong EoS \cite{RedlichKwong1949}, cannot explicitly provide model parameters that can be used to maintain pressure equilibrium. This makes oscillation-free schemes designed for specific EoS difficult to directly apply to more general real fluids.

Despite these difficulties, over the past decade many oscillation-free schemes have been proposed for real fluids with arbitrary equations of state. In general, these schemes follow the second type of the classical two-phase-flow methods discussed above, assuming that the fluids within a finite-volume cell reach an instantaneous pressure equilibrium, while this equilibrium is maintained by modifying the energy equation or the flux construction. The most direct approach is the pressure-based (PB) method \cite{terashima2012,Kawai2015,Kitamura2018}, in which the energy equation is replaced by the pressure equation. Similar methods extended from the PB method include the pressure-equation-projected DG method \cite{Ching2025}, the conservative/non-conservative adaptive method \cite{Xu2026}, and the enthalpy-based method \cite{Lacaze2019}. Other methods modify the energy flux, including the approximating equilibrium scheme \cite{terashima2025} and the double-flux (DF) method \cite{Ma2017,rodriguez2018,yatsuyanagi2022,Zhang2024}. Notably, the DF method introduces a "frozen" concept to extend the classic multi-phase double-flux approach to real fluids. Due to its simple and natural construction, it remains the most widely used method in the field of real-fluid simulations. However, when confronting more extreme phase-change flow problems, the DF method often exhibits compromised accuracy and stability.

In our recent work, we proposed a Real Fluid Quasi-Conservative (RFQC) method \cite{Bai2026RFQC}. Unlike the above methods, the RFQC method does not modify the form of the energy equation or the energy flux. Instead, it introduces additional thermodynamic advection equations and evolves them alongside the Euler equations:
\begin{equation}
\begin{dcases}
\frac{\partial \mathbf{U}}{\partial t} + \frac{\partial \mathbf{F}}{\partial x} =0, \\
\frac{\partial \boldsymbol{\Phi}}{\partial t} + u \frac{\partial \boldsymbol{\Phi}}{\partial x} = 0.
\end{dcases}
\end{equation}
where $\mathbf{U} = (\rho \quad \rho u \quad \rho e_t)^T$ denotes the conservative variables, $\mathbf{F} = (\rho u \quad \rho u^2 + p \quad (\rho e_t + p)u)^T$ denotes the flux vector, and $\boldsymbol{\Phi} = (\xi \quad E_0)^T$ denotes the thermodynamic affine variables. The thermodynamic variables are defined as: 
\begin{equation}
\xi = \frac{h}{c^2} = \frac{e+p/\rho}{c^2},\quad E_0 = \rho e - \xi p .
\end{equation}
The oscillation-free pressure is then recovered from the conservative variables and the thermodynamic variables:
\begin{equation}
p = \frac{\rho e - E_0}{\xi}.
\end{equation}
Finally, the RFQC method performs a thermodynamic re-projection. The updated density and the recovered pressure $(\rho,p)$ are combined with the EoS to recalculate the thermodynamic state $(\rho e,\xi,E_0)$ for the next time step. This ensures the correct evolution of $\boldsymbol{\Phi}$ and satisfies the Abgrall condition. In various real-fluid phase-change Riemann problems \cite{Bai2026FeRP}, the RFQC method shows higher accuracy than the PB method and the DF method.

Although oscillation-free schemes for real fluids have made significant progress and have advanced from the initial transcritical problems to increasingly complex phase-change problems, a current limitation of the real-fluid oscillation-free schemes should still be pointed out by comparing their development with that of classical two-phase-flow schemes. Unlike classical two-phase methods that possess a clear equilibrium relaxation mechanism, the development of oscillation-free schemes for real fluids has been primarily method-driven, with their underlying physical understanding remaining under-explored. However, advancements in physical understanding are indispensable for the rational design and long-term development of numerical methods. It is worth emphasizing that, whether in the pressure-based method, the enthalpy-based method, the double-flux method, or the RFQC method, the motivation is to resolve the conflict between the thermodynamic nonlinearity of real fluids and the averaging of conservative variables, rather than to answer how mechanical equilibrium of real fluids should be physically recovered. Therefore, in Section 2, from the perspective of continuum thermodynamics, we analyze how the traditional finite-volume method recovers pressure through the thermodynamic equilibrium assumption, and how this thermodynamic equilibrium violates pressure equilibrium in real fluids. Furthermore, Section 3 interprets how the RFQC method physically recovers the oscillation-free pressure from the perspective of providing mechanical-equilibrium information to real-fluid elements, and clarifies the physical origins of its thermodynamic-variable evolution and re-projection errors.

Additionally, real fluids experience an abrupt jump in the speed of sound during phase change. In the single-phase region, the speed of sound is on the order of $O(10^2)$--$O(10^3)$, whereas inside the two-phase mixture region, it drops to only $O(10)$ or even $O(1)$. The speed of sound in the two-phase region can be written as
\begin{equation}
\frac{1}{\rho c_{eq}^2}=\frac{\alpha }{{\rho_v}c_v^2}+\frac{1-\alpha }{{\rho_l}c_l^2}+T\left( \frac{\alpha {\rho_v}}{{C_{p,v}}}{{\left( \frac{\text{d}s_v}{\text{d}p} \right)}^{2}}+\frac{(1-\alpha ){\rho_l}}{C_{p,l}}{{\left( \frac{\text{d}s_l}{\text{d}p} \right)}^{2}} \right)
\label{equ:c_eq}
\end{equation}
where $\alpha$ is the vapor volume fraction, and $c_v$, $c_l$ and $C_{p,v}$, $C_{p,l}$ denote the speeds of sound and specific heats at constant pressure at $(p,T)$ on the saturation line, respectively. The term in the parentheses on the right-hand side represents the equilibrium heat and mass transfer between the two phases, which is the source of the speed of sound discontinuity \cite{Bai2026FeRP}.

This abrupt jump in the speed of sound poses a challenge to current real-fluid schemes. Not only do the double-flux method and the pressure-based method tend to produce large errors in such problems \cite{Bai2026RFQC}, but we also find that the RFQC method exhibits a numerical anomaly in a special class of phase-change Riemann problems with liquid-upwind advection, where the rarefaction wave cannot develop correctly. For this reason, in Sections \ref{sec:LUA_char} and \ref{sec:LUA_mec}, we introduce and analyze this numerical anomaly, and in Section \ref{sec:LUA_reg} we give a remedy based on regularized initial conditions. Finally, the robustness of the RFQC method in such challenging phase-change problems is validated through two-dimensional sonic phase-change jet cases.

\section{What is Pressure Oscillation in Physics}
We point out that, from a physical viewpoint, this problem can be formulated as how pressure should be recovered from a finite-volume cell. The traditional finite-volume method recovers pressure based on the thermodynamic equilibrium assumption, whereas oscillation-free schemes recover pressure through the mechanical-equilibrium assumption.

\subsection{Finite-volume update}
We first consider the energy conservation equation of the one-dimensional Euler equations. At a contact discontinuity where the velocity $u=u_0$ and the pressure $p=p_0$ are both constant, the kinetic-energy term and the pressure-work term cancel each other, and the energy equation reduces to a pure advection equation for the volumetric internal energy $\rho e$:
\begin{equation}
\frac{\partial (\rho e)}{\partial t} + u_0 \frac{\partial (\rho e)}{\partial x} = 0
\label{equ}
\end{equation}
Consider a first-order finite-volume scheme. During the update from time step  $n$ to $n+1$, the conservative variables $U$ in a cell are the weighted average of the fluid elements that enter this cell during the evolution. Suppose that two fluid elements enter this cell, denoted by the subscripts $L$ and $R$. Let $\lambda$ be the weighting coefficient determined by the CFL condition and the advection velocity. The discrete update can then be written as:

\begin{equation}
\begin{aligned}
\rho_i^{n+1} &= (1-\lambda)\rho_L + \lambda \rho_R = \bar{\rho} \\
(\rho e)_i^{n+1} &= (1-\lambda)(\rho e)_L + \lambda (\rho e)_R = \overline{\rho e}
\end{aligned}
\end{equation}

\begin{figure}[h]
\centering 
\includegraphics[width=0.5\textwidth]{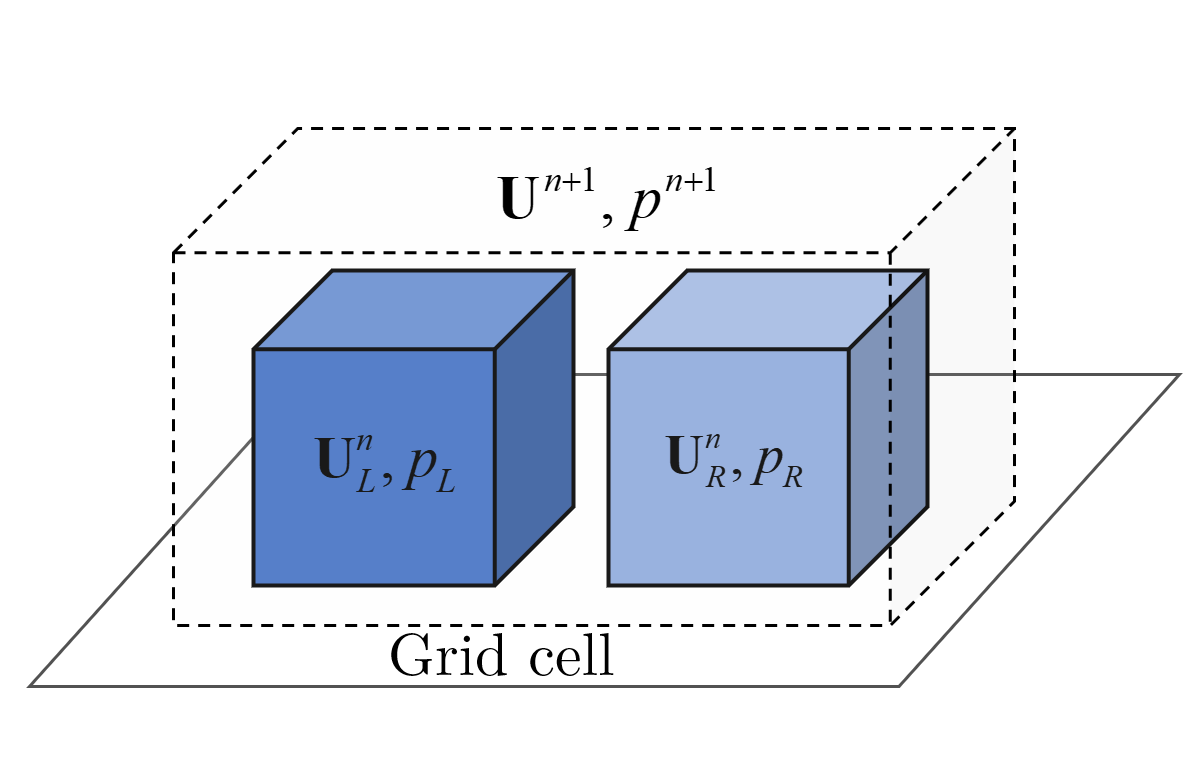} 
\caption{Diagram of the finite-volume update.}
\label{fig1}
\end{figure}

At time $n+1$, the state inside the finite-volume cell is shown in Fig.~ \ref{fig1}. What is known as the pressure oscillation problem refers to how we should resolve the pressure within this finite-volume cell.

\begin{itemize}
\item \textbf{Thermodynamic equilibrium perspective}: In traditional finite-volume methods, the pressure is directly recovered from the conservative variables through
$p^{n+1} =p\left( \bar{\rho}, \frac{\overline{\rho e}}{\bar{\rho}} \right)= p\left( \bar{\rho}, \bar{e} \right) $. The underlying logic of this treatment comes from thermodynamic equilibrium: when a continuum element is in thermodynamic equilibrium, its macroscopic thermodynamic state is completely determined by a thermodynamic potential formed by two independent thermodynamic parameters \cite{Sedov1973}. The internal energy, density, and entropy form the thermodynamic potential $e = e(\rho,s)$. Meanwhile, since the thermodynamic absolute temperature satisfies $T=(\partial e/\partial s)_\rho>0$, the inverse function $s = s(e, \rho)$ exists uniquely. Therefore, solely based on $(\bar{\rho}, \bar{e})$, the entire macroscopic thermodynamic state of the fluid element, including the pressure $p$, can be uniquely recovered. In this sense, recovering pressure with a two-parameter EoS in the traditional finite-volume method is equivalent to instantaneously homogenizing different fluid elements in the cell into a thermodynamic equilibrium final state.
\item \textbf{Mechanical equilibrium perspective}: In a finite-volume cell that has achieved pressure equilibrium, the pressure update must continue to satisfy the mechanical equilibrium condition $p^{n+1} = p_0 $. The underlying logic originates from the Euler equations: pressure changes should result only from volumetric work rather than thermal effects. Therefore, to obtain the mechanical-equilibrium pressure $p$ from the finite-volume cell at time $n+1$, additional information representing mechanical equilibrium is required besides $(\bar{\rho}, \bar{e})$.
\end{itemize}

\subsection{Physical explanation of pressure oscillations}

Why are traditional finite-volume methods free of pressure oscillations when calculating ideal gases? We point out that this is because the thermodynamic equilibrium process for an ideal gas happens to be a strictly isobaric process. For an ideal gas, since there is no molecular potential energy, the pressure and the volumetric internal energy have a linear relation, and both originate only from the thermal motion of molecules, namely
\begin{equation}
p = (\gamma - 1) \rho e = \frac{1}{3} n m \overline{v^2}
\end{equation}
where $n$ is the number density, $m$ is the molecular mass, $v$ is the thermal motion velocity, and $\gamma$ is the specific heat ratio of the ideal gas. Therefore, the isochoric-adiabatic mixing process of an ideal gas in a control volume is a strictly isobaric process. Even if the traditional finite-volume method recovers pressure from the thermodynamic equilibrium perspective, its mechanical equilibrium condition is still not violated:
\begin{equation}
p^{n+1} = (\gamma - 1) \overline{\rho e} = \bar p = p_0
\end{equation}

However, for real fluids, molecular potential energy also exists between fluid molecules, and the linear relation between internal energy and pressure is generally not satisfied. The isochoric-adiabatic mixing process in a control volume is therefore not isobaric. If these fluid elements are forced to reach thermodynamic equilibrium in a finite-volume cell, the pressure changes, namely $p(\bar{\rho}, \bar{e}) \neq p_0$. This underpins the widely used explanation of "the conflict between a nonlinear EoS and linear averaging \cite{Bai2026RFQC}."

We emphasize that this mechanism represents the physical origin of pressure oscillations; it is fundamentally a matter of choosing the appropriate equilibrium state, rather than a closure issue of the energy equation. If the Godunov-method perspective is adopted and the pressure is updated through the thermodynamic equilibrium assumption, pressure oscillations will occur inevitably. However, the thermodynamic equilibrium assumption is not always acceptable. Constrained by complex real-fluid EoS, the recovered pressure can easily drop into negative regimes, leading to numerical breakdown. In real-fluid CFD simulations, what we truly require is a flow field that satisfies mechanical equilibrium from the perspective of the Euler equations.

\section{How to Recover Pressure From a Finite-volume Averaged Cell}

As pointed out previously, calculating pressure by combining $(\bar{\rho}, \bar{e})$ with the EoS implicitly introduces the thermodynamic equilibrium assumption. To calculate the pressure $p$ from the perspective of mechanical equilibrium, we should provide additional information regarding mechanical equilibrium. How then do we find this information? We need to consider two constraints: first, this information must satisfy the isentropic constraint; second, it must be amenable to finite-volume averaging. In the following, we will construct the pressure recovery formulation from the affine relationship between internal energy and pressure under isentropic conditions, and examine how the RFQC algorithm realizes this operation.

\subsection{Affine relationship between internal energy and pressure}

In the smooth regions of the Euler equations, entropy is the sole invariant carried by a fluid element, satisfying:
\begin{equation}
\frac{Ds}{Dt}=0.
\label{eq:ds}
\end{equation}

Under the isentropic condition, we consider an affine form of the non-linear relation between internal energy and pressure:
\begin{equation}
\rho e = \xi p + E_0.
\label{eq:rhoe_xi}
\end{equation}
Here, $\xi$ represents the isentropic response of internal energy to pressure, i.e., the linear coefficient of the affine relationship. It indicates the increment of internal energy absorbed per unit volume of fluid for a unit increase in pressure under isentropic compression. 
$E_0$ denotes the reference internal energy density at a "zero-pressure" state, which encompasses background energy such as the intermolecular cohesive potential energy of the fluid. From the fundamental thermodynamic relation $\mathrm{d}e = T \mathrm{d}s + {p}/{\rho^2} \mathrm{d}\rho$, we obtain:

\begin{equation}
\xi = \pderiv{(\rho e)}{p}{s} = \frac{e + {p}/{\rho}}{c^2} = \frac{h}{c^2}
    \label{eq:xi}
\end{equation}

Meanwhile, $\xi$ and $E_0$ are both quantities that can be finite-volume averaged, which is equivalent to the mechanical equilibrium condition. Suppose that several fluid elements with different thermodynamic states exist in a finite-volume cell, with the element index denoted by $j$. According to the mechanical equilibrium assumption, these elements instantaneously reach pressure equilibrium inside the cell, namely they share the same macroscopic pressure $p$, without requiring thermodynamic equilibrium. When performing a finite-volume average on the total internal energy within the cell (with weights $\lambda_j$), the shared pressure $p$ can be extracted from the summation symbol as a constant:

\begin{equation}
    \overline{\rho e} = \sum \lambda_j \left( p \xi_j +E_{0,j}\right)  =p \left( \sum \lambda_j \xi_j \right) + \sum \lambda_j E_{0,j} = \bar{\xi} p + \bar{E_0}
\end{equation}

\begin{figure}[h]
\centering 
\includegraphics[width=0.4\textwidth]{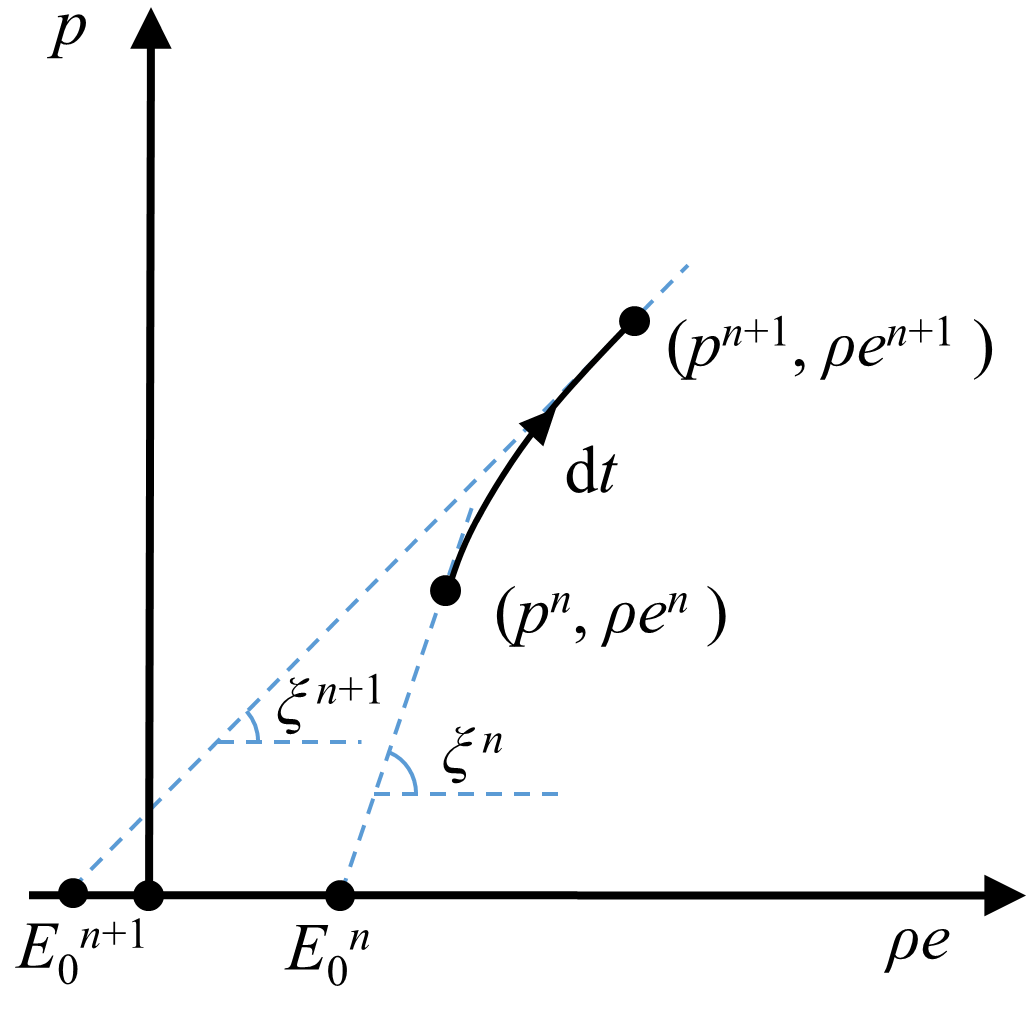} 
\caption{Isentropic pressure evolution.}
\label{fig2}
\end{figure}

Figure~\ref{fig2} shows the internal-energy--pressure evolution curve of a single fluid element under the isentropic condition. During the time interval $\mathrm{d}t$, the fluid element evolves from state $n$ to state $n+1$. Therefore, theoretically, if $\xi$ and $E_0$ are evolved synchronously in the finite-volume update, the mechanical-equilibrium pressure $p^{n+1}$ in the cell can be recovered from the cell-averaged conservative energy $\overline{\rho e} = \sum \lambda_j(\rho e)^{n+1}_j$ and the affine coefficients $\bar{\xi} = \sum \lambda_j\xi^{n+1}j,\bar{E_0} = \sum \lambda_jE^{n+1}{0,j}$:

\begin{equation}
p^{n+1}=\frac{\overline{\rho e} - \bar{E_0}}{\bar{\xi}}
\end{equation}

We next derive the evolution equations of $\xi$ and $E_0$ from the isentropic relation.

We choose $(\rho, s)$ as the state parameters, so that $\xi = \xi(\rho, s)$. The material derivative of $\xi$ along a fluid element is
\begin{equation}
    \sderiv{\xi}{t} = \pderiv{\xi}{\rho}{s}\sderiv{\rho}{t}+\pderiv{\xi}{s}.{\rho}\sderiv{s}{t}
\end{equation}
Substituting the continuity equation $\mathrm{D}\rho /\mathrm{D}t = -\rho \partial u/ \partial x$ and the isentropic condition (Eq. ~\ref{eq:ds}) yields:
\begin{equation}
    \sderiv{\xi}{t} = -\rho \pderiv{\xi}{\rho}{s} \frac{\partial u}{\partial x}.
\end{equation}
To express this result in terms of the pressure--internal-energy response, the chain rule under the isentropic condition is used to replace $\rho$ by $p$:
\begin{equation}
\sderiv{\xi}{t} = -\rho c^2\pderiv{\xi}{p}{s}\frac{\partial u}{\partial x}
\label{eq:Dxi}
\end{equation}
Its Eulerian form is given by:
\begin{equation}
    \frac{\partial \xi}{\partial t} + u \frac{\partial \xi}{\partial x} = -\rho c^2\pderiv{\xi}{p}{s}\frac{\partial u}{\partial x}
    \label{eq:pxi}
\end{equation}

The derivation for $E_0$ is similar. From the definition $E_0=\rho e-\xi p$, taking the material derivative of $E_0$ gives:
\begin{equation}
    \sderiv{E_0}{t} = \sderiv{(\rho e)}{t} - \xi\sderiv{p}{t}- p\sderiv{\xi}{t}.
    \label{eq:DE_0}
\end{equation}

On the other hand, from $\rho e=(\rho e)(p,s)$, we have:
\begin{equation}
    \sderiv{(\rho e)}{t} = \pderiv{(\rho e)}{p}{s} \sderiv{p}{t} +   \pderiv{(\rho e)}{s}{p} \sderiv{s}{t}
\end{equation}
Substituting the definition of $\xi$ (Eq. ~\ref{eq:xi}) and the isentropic condition  (Eq. ~\ref{eq:ds}) yields:
\begin{equation}
    \sderiv{(\rho e)}{t} = \xi \sderiv{p}{t} 
    \label{eq:Drhoe1}
\end{equation}
Substituting Eq.~\ref{eq:Drhoe1} and Eq.~\ref{eq:Dxi} back into Eq.~\ref{eq:DE_0} provides the evolution equation for $E_0$:
\begin{equation}
\sderiv{E_0}{t} =  p\rho c^2\pderiv{\xi}{p}{s}\frac{\partial u}{\partial x}
\end{equation}

Its Eulerian form is:
\begin{equation}
    \frac{\partial E_0}{\partial t} + u \frac{\partial E_0}{\partial x} =  p\rho c^2\pderiv{\xi}{p}{s}\frac{\partial u}{\partial x}
\end{equation}

In summary, the evolution equations for the slope $\xi$ and intercept $E_0$ of the internal energy--pressure affine relation are:
\begin{equation}
\begin{dcases}
    \frac{\partial \xi}{\partial t} + u \frac{\partial \xi}{\partial x} = -\rho c^2\pderiv{\xi}{p}{s}\frac{\partial u}{\partial x} \\
    \frac{\partial E_0}{\partial t} + u \frac{\partial E_0}{\partial x} =  p\rho c^2\pderiv{\xi}{p}{s}\frac{\partial u}{\partial x}
    \label{eq:p_xi1}
\end{dcases}
\end{equation}
where the isentropic derivative of $\xi$ with respect to $p$ is a quantity related to the second-order curvature of the isentrope (i.e., the fundamental derivative of gasdynamics):
\begin{equation}
    \pderiv{\xi}{p}{s}=\frac{1}{\rho c^2}-\frac{\xi}{c^2}\pderiv{c^2}{p}{s}
\end{equation}

\subsection{What does RFQC do}

For certain specific equations of state, such as the van der Waals EoS and the stiffened-gas EoS, the internal-energy--pressure affine relation discussed above can be written analytically, and its evolution equations can be constructed in a relatively direct way \cite{Shyue1998,Pantano2017}. For instance, the affine coefficients of the stiffened gas EoS are all constants; thus, their source terms are naturally zero, and Eq.~\ref{eq:p_xi1} degenerates into advection equations for $\xi$ and $E_0$. However, for more complex real-fluid equations of state, such as cubic EoS and composite EoS under two-phase equilibrium, the internal energy--pressure affine relationship cannot be directly inverted from the EoS. Meanwhile, the source terms involving the derivative of the speed of sound are even more difficult to calculate. To the best of our knowledge, within the multi-component HEM framework, there is still no mature analytical framework for calculating the derivatives of the two-phase equilibrium speed of sound.

The RFQC method is proposed for this class of real fluids. We recall the transport equations for the thermodynamic affine variables $\boldsymbol{\Phi}=(\xi, E_0)^T$:
\begin{equation}
\begin{dcases}
    \frac{\partial \xi}{\partial t} + u \frac{\partial \xi}{\partial x} = 0 \\
    \frac{\partial E_0}{\partial t} + u \frac{\partial E_0}{\partial x} = 0
    \label{eq:p_xi_RFQC}
\end{dcases}
\end{equation}

Comparing Eq.~\ref{eq:p_xi1} with Eq.~\ref{eq:p_xi_RFQC}, we can see that the RFQC method truncates the source terms in the isentropic evolution equations of $\xi$ and $E_0$. As a result, $\xi$ and $E_0$ are evolved along material trajectories (pathlines) rather than along isentropic trajectories. We point out that this is the key point in the construction of the RFQC method: on the one hand, RFQC avoids pressure oscillations by evolving the affine coefficients of the internal energy--pressure relationship; on the other hand, by freezing $\xi$ and $E_0$ along pathlines, it avoids the evaluation of source terms related to the derivatives of the speed of sound, while the re-projection converts the accumulated error caused by the off-isentropic evolution of $\xi$ and $E_0$ into an internal-energy residual, through which the thermodynamic state is re-anchored rigorously at the next time step. 

\begin{figure}[htbp]
\centering
\includegraphics[width=0.95\textwidth]{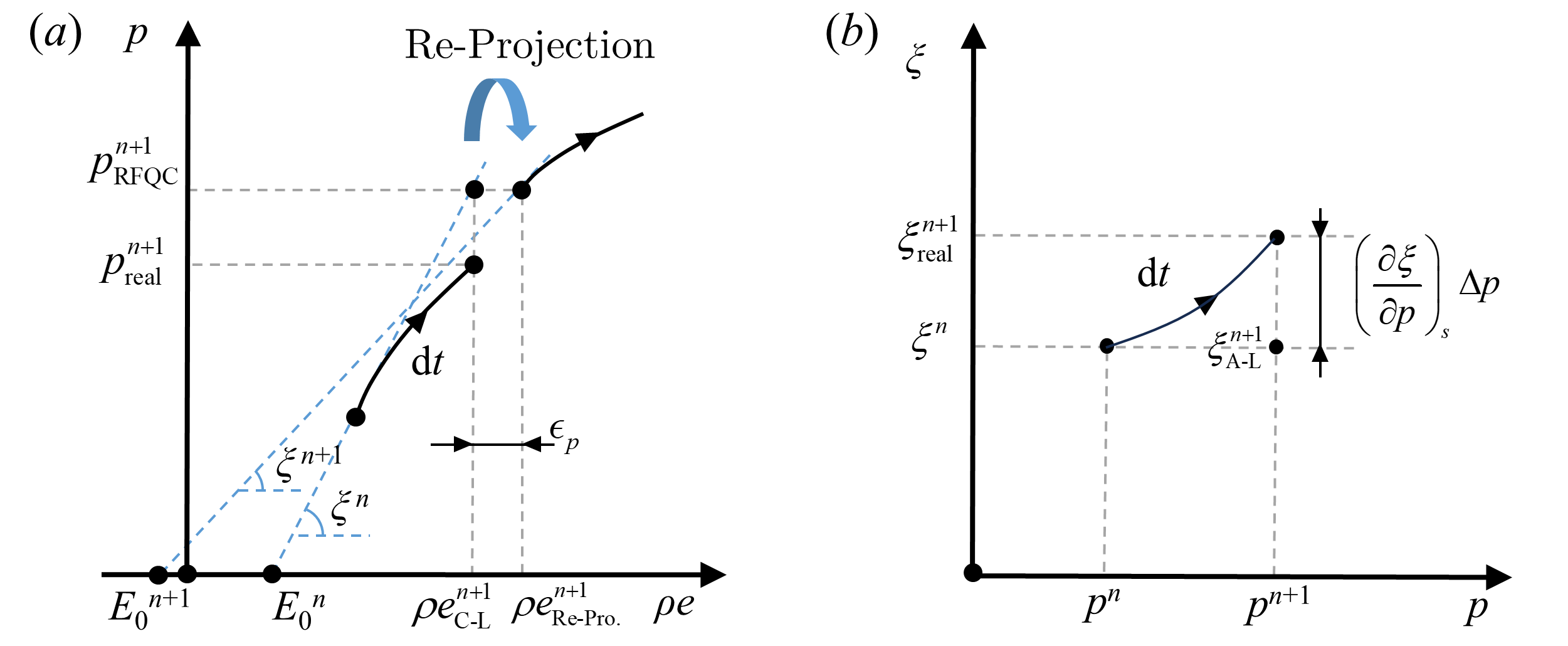}
\caption{(a) Pressure evolution; (b) Affine coefficient evolution. Diagram of the RFQC algorithm on an isentropic surface.}
\label{fig3}
\end{figure}

Figure~\ref{fig3} shows the evolution trajectories of pressure and the affine coefficient $\xi$ on an isentropic surface in the RFQC algorithm. It provides a more direct interpretation of the physical meaning of "internal-energy--pressure linearization--frozen evolution of the linearization coefficients--pressure-field reconstruction--thermodynamic re-projection" in the RFQC algorithm.

\begin{itemize}
\item \textbf{Internal-energy--pressure linearization:} From the physical perspective, the local linearization of the internal-energy--pressure relation in RFQC is, fundamentally, the exact local affine relation between internal energy and pressure. Mechanical equilibrium of pressure is maintained through the affine-relation coefficients that can be finite-volume averaged.

\item \textbf{Frozen evolution of the linear coefficients:} The RFQC algorithm assumes that the affine coefficients $\xi^n,E_0^n$ are frozen along pathlines and transported by pure advection. This is equivalent to approximating the internal-energy--pressure isentrope $\rho e = \psi(p,s)$ within the time interval $\mathrm{d}t$ by the tangent line at time $t^n$ (Fig.~\ref{fig3}(a)).

\item \textbf{Pressure-field reconstruction:} The internal energy $\rho e^{n+1}_\text{C-L}$ evolved by the conservation laws and the affine coefficients $\xi^{n+1}_\text{A-L}=\xi^n,E^{n+1}_\text{0,A-L} = E_0^n$ evolved by advection are used to recover the mechanical-equilibrium pressure $p^{n+1}_\text{RFQC}$. There is an error between $p^{n+1}_\text{RFQC}$ and the actual isentropic evolution pressure of the fluid element, $p^{n+1}_\text{real}$. This error originates from the second-order curvature term of $\psi(p,s)$, and its sign depends on the convexity of $\psi$.

\item \textbf{Thermodynamic re-projection:} The pressure $p^{n+1}_\text{RFQC}$ and the density $\rho^{n+1}$ are then used as the thermodynamic reference parameters at time $t^{n+1}$ to recalculate the thermodynamic state at time $t^{n+1}$, including the internal energy $\rho e^{n+1}$ and the affine coefficients $\xi^{n+1},E^{n+1}$. The difference $\epsilon_p$ between $\rho e^{n+1}$ and $\rho e^{n+1}_\text{C-L}$ is the re-projection error, which also depends on the second-order curvature term $\pderiv{\xi}{p}{s}(\Delta p)^2$ of $\psi(p,s)$.

\end{itemize}

The present analysis demonstrates that by truncating the source terms in the isentropic evolution equations of the affine coefficients (Eq.~\ref{eq:p_xi1}), the RFQC method strikes a favorable balance among computational efficiency, shock-capturing stability, and conservation errors. Nevertheless, the theoretical framework established above also provides guidance for designing non-oscillatory schemes of even higher accuracy, such as incorporating the source terms into the evolution or recovering the equilibrium pressure via more complex internal energy–pressure relations (e.g., a quadratic relation). Predictably, however, such advancements will necessitate the evaluation of more complex thermodynamic derivatives and pose challenges for numerical stability.

\section{Applicability Limit of RFQC: the Liquid-upwind Anomaly}

\label{sec:LUA_char}
When the RFQC algorithm is applied to solve a Riemann problem involving phase change from a high-pressure liquid to a low-pressure vapor, and an additional liquid-upwind translational velocity is superimposed on the initial conditions (the liquid phase is on the upstream side), we observe a highly reproducible anomaly. We designate this anomaly as the Liquid-upwind Anomaly (LUA). Below, we discuss the behavior of this anomaly systematically and analyze its origin. Throughout the subsequent analysis, we consider the RFQC method with first-order temporal and spatial discretization, utilizing the HLLC Riemann solver \cite{Toro2009}. The fluid medium under consideration is n-dodecane. For detailed information regarding the discretization schemes, definitions of thermodynamic variables, equations of state, and phase equilibrium solvers, we refer readers to our previous work \cite{Bai2026RFQC} and omit the details here.

Figure~\ref{fig4} shows a typical case of LUA. The computational domain is $L=1\mathrm{m}$ and the initial discontinuity is located at $x=0.5\mathrm{m}$. The initial states are $p_L = 2 \times 10^6 \mathrm{Pa}$, $p_R = 1 \times 10^5\mathrm{Pa}$, $\rho_L = 300\mathrm{kg/m^{3}}$, and $\rho_R = 2 \mathrm{kg/m^{3}}$. This represents a class of Riemann problems involving liquid--vapor phase change, also known as flash evaporation Riemann problems. The exact solution of this type of Riemann problem was given in our previous work \cite{Bai2026FeRP}. The subsequent discussion in Sections \ref{sec:LUA_char} to \ref{sec:LUA_reg} uses this case as the representative example. In Fig.~\ref{fig4}, as the global translational velocity changes, the numerical solution obtained by the RFQC method deviates significantly from the exact solution. The main characteristics are as follows:

\begin{figure}[htbp]
\centering
\includegraphics[width=0.95\textwidth]{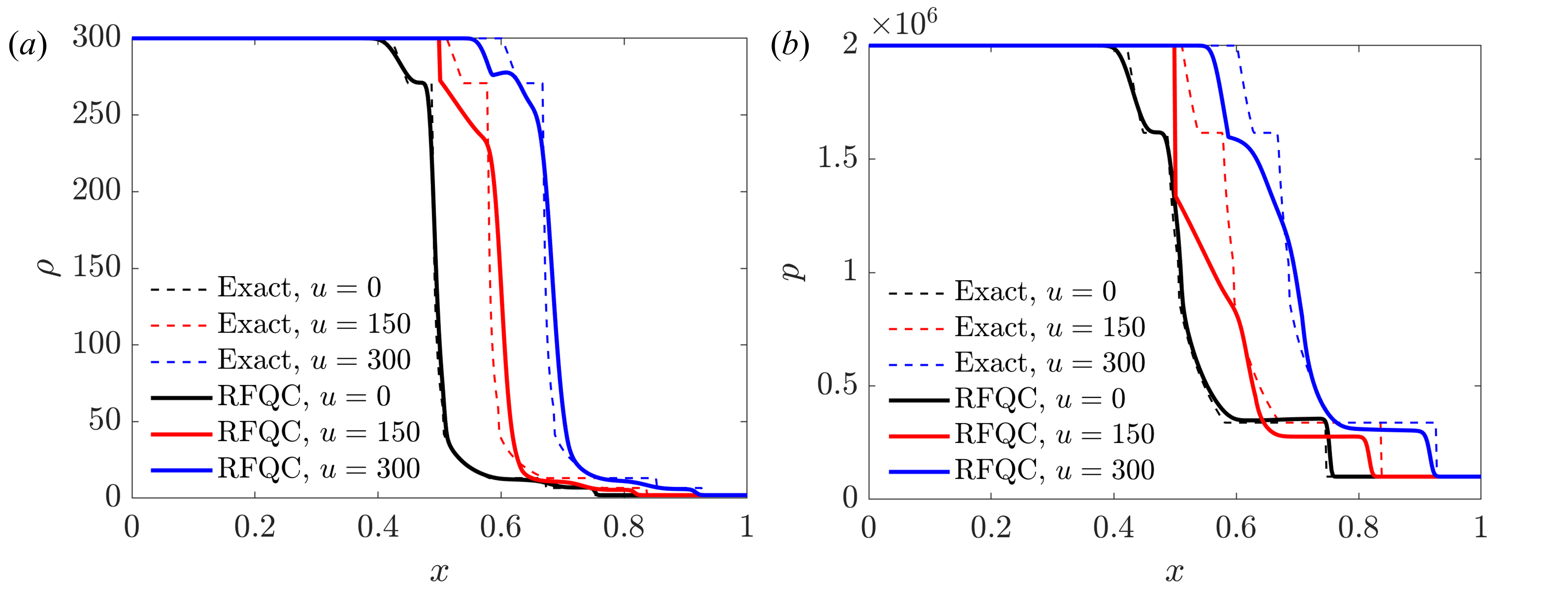} 
\caption{(a) Density; (b) Pressure. Liquid-upwind anomaly in the RFQC method. The computational time $t=0.6\mathrm{ms}$, with a grid size of $N=500$ and a CFL number of $0.5$.}
\label{fig4}
\end{figure}

\begin{figure}[htbp]
\centering
\includegraphics[width=0.95\textwidth]{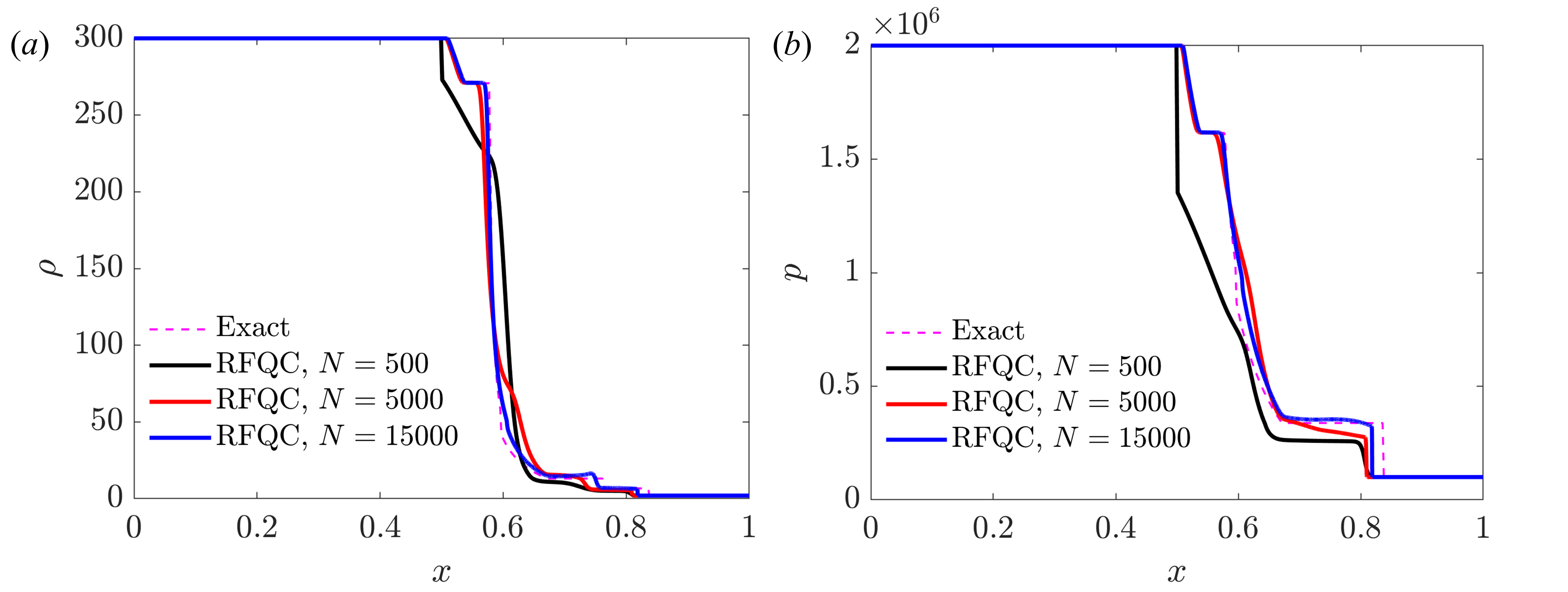}
\includegraphics[width=0.95\textwidth]{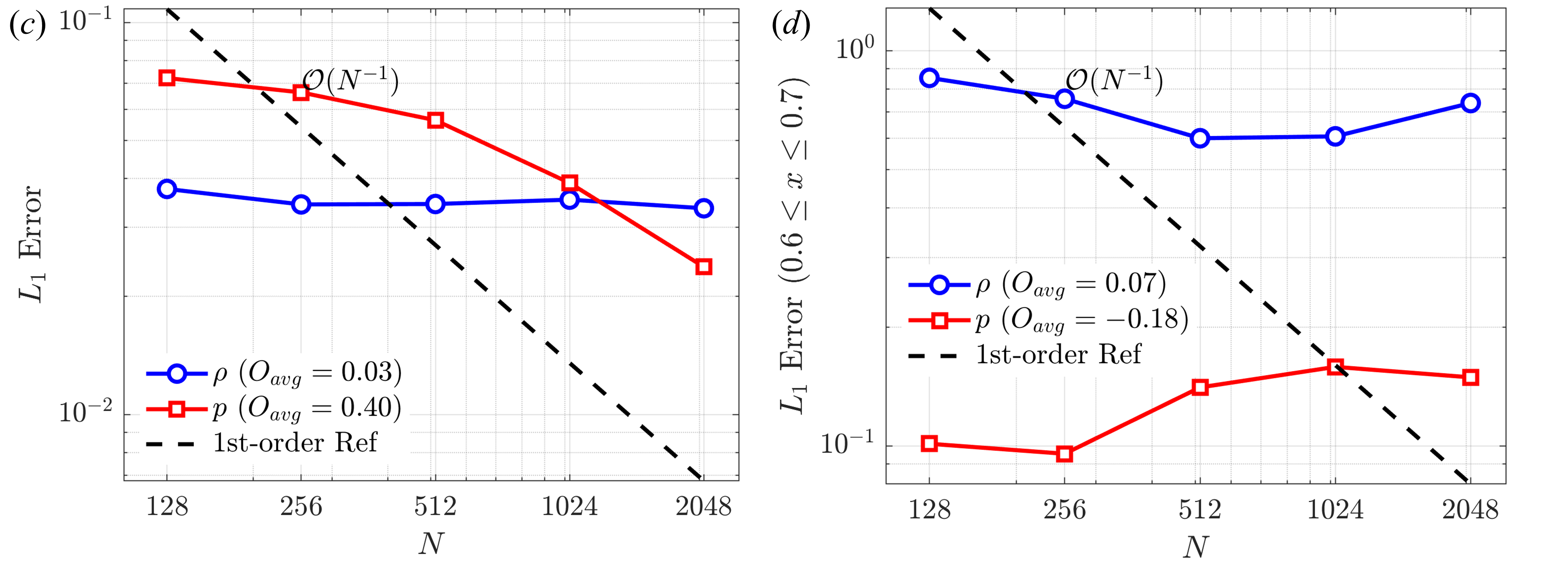}
\caption{(a) Density; (b) Pressure; (c) Global $L_1$ error; (d) $L_1$ error in smooth region. Grid-convergence behavior of the liquid-upwind anomaly. The computational time $t=0.6\mathrm{ms}$, initial translational velocity $u =150\mathrm{m/s}$, $\text{CFL} =0.5$.}
\label{fig5}
\end{figure}

\begin{figure}[htbp]
\centering
\includegraphics[width=0.95\textwidth]{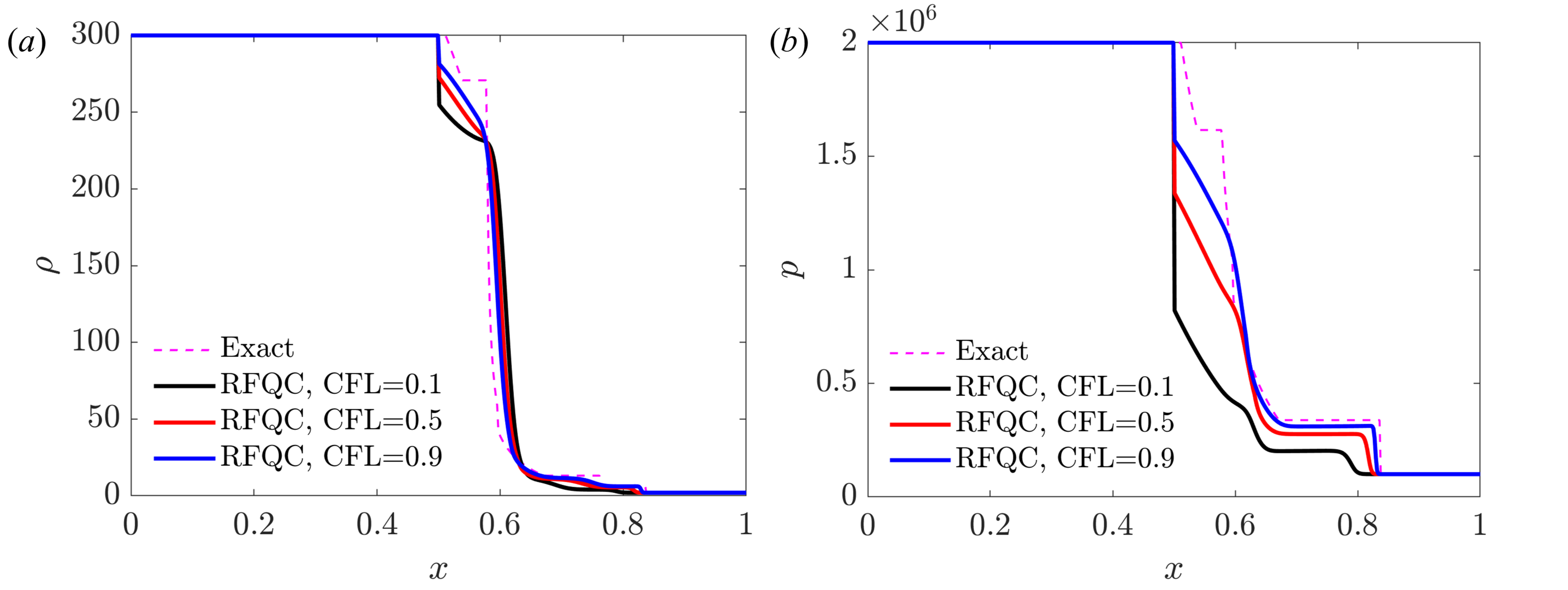}
\caption{(a) Density; (b) Pressure. Effect of the CFL number on the liquid-upwind anomaly. The computational time $t=0.6\mathrm{ms}$, initial translational velocity $u =150\mathrm{m/s}$, grid size $N=500$.}
\label{fig6}
\end{figure}

\begin{itemize}
    \item \textbf{Residual initial discontinuity:} The initial discontinuity remains over many time steps, so that the rarefaction fan cannot develop correctly and the downstream pressure stays at low values. This behavior is observed near $x=0.5\mathrm{m}$ for the $u=150\mathrm{m/s}$ curve in Fig.~\ref{fig4}.
    \item \textbf{Non-monotonic density inside the rarefaction wave:} A local density rebound appears inside the rarefaction fan, where the density should decrease monotonically, or the curvature of the profile changes significantly. This can be seen near $x=0.65\mathrm{m}$ for the $u=300\mathrm{m/s}$ curve in Fig.~\ref{fig4} (a).
    \item \textbf{Grid convergence:} The wave profile improves clearly with grid refinement, as shown in Figs.~\ref{fig5}(a) and \ref{fig5}(b). However, the error convergence order remains very low, as shown in Figs.~\ref{fig5}(c) and \ref{fig5}(d).
    \item \textbf{CFL number:} Increasing the CFL number slightly improves the wave profile, as shown in Fig.~\ref{fig6}, but the solution cannot be restored to the correct state observed at $u=0$ in Fig.~\ref{fig4}.
 \item \textbf{Insensitivity to time integration:} This anomaly is generally insensitive to the time-integration method, with similar results obtained from first-, second-, and third-order Runge--Kutta schemes.
\end{itemize}

The above anomaly usually occurs only when the following three conditions are satisfied simultaneously:

\begin{itemize}
\item \textbf{Splitting of the rarefaction wave:} The rarefaction wave is split by phase change, and the speed of sound changes discontinuously at the end of the split plateau due to phase change. The speed of sound inside the two-phase region becomes very low, only on the order of $O(1)$--$O(10)$.

\item \textbf{Global advection:} A significant global translational velocity is prescribed in the initial condition, rather than superimposed later.

\item \textbf{Liquid upwind:} The upstream state is a high-pressure, high-density liquid, while the downstream state is a low-pressure (below the critical pressure), low-density vapor/two-phase state.
\end{itemize}

Unlike the well-known shock anomaly, the triggering condition of the LUA does not require strong compression in the flow field; instead, the LUA appears specifically in rarefaction-wave and phase-change problems.

\section{Formation Mechanism of the Liquid-Upwind Anomaly}
\label{sec:LUA_mec}
\begin{figure}[htbp]
\centering
\includegraphics[width=0.55\textwidth]{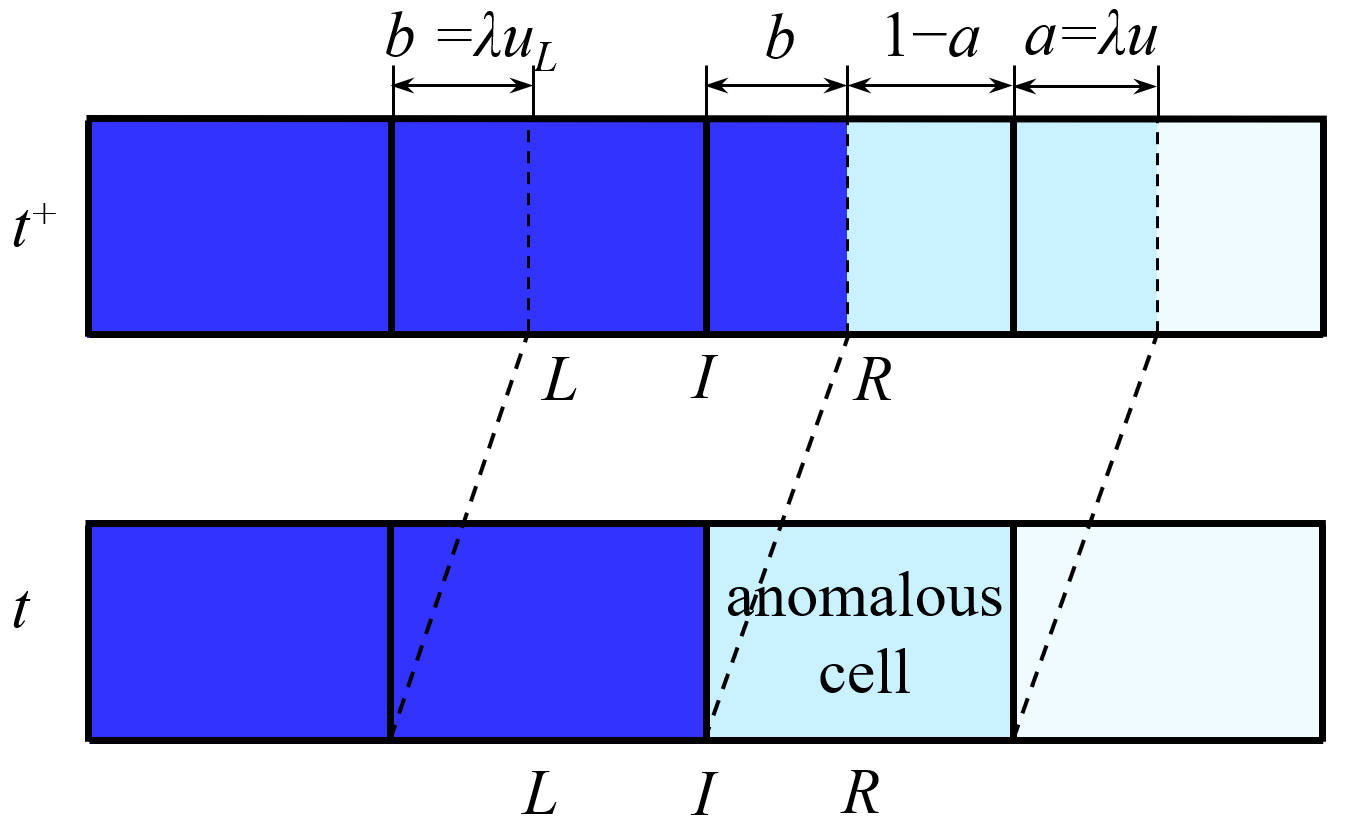}
\caption{Diagram of the cells on both sides of the initial discontinuity.}
\label{fig7}
\end{figure}

We now elucidate the formation mechanism of the LUA and interpret the reasons for the distinct behaviors detailed above. As shown in Fig.~\ref{fig7}, we consider the numerical update of the two grid cells adjacent to the initial discontinuity in the RFQC algorithm. The cell to the left of the initial discontinuity interface $I$ is denoted by $L$, and the cell to the right is denoted by $R$, with the interface located at $I=L+1/2=R-1/2$. Initially, cell $L$ contains high-pressure liquid, whereas the $R$ cell contains low-pressure vapor. At time $t$, cell $R$ shifts into a two-phase mixed state due to an increase in both pressure and density.

We first point out that the root cause of the LUA lies in the delayed pressure rise in cell $R$, which prevents a continuous rarefaction wave from forming in time. In a regular evolution scenario under a liquid-upwind velocity, the initial discontinuity interface $I$ is expected to be located within or upstream of the rarefaction fan. The downstream $R$ cell should firstly be lifted toward the intermediate pressure of the Riemann solution, gradually rising to connect with the left state to form a continuous wave fan, or be raised to the upstream pressure, becoming part of the liquid region upstream of the rarefaction wave. In the RFQC method, however, once the $R$ cell enters the two-phase region during phase change, the rapid drop in the speed of sound leads to a sharp increase in the affine slope $\xi$. The pressure increment driven by the upstream liquid flux is therefore very limited. Meanwhile, the thermodynamic re-projection performed at each time step equivalently removes the local positive pressure increment. The pressure rise in the $R$ cell is consequently delayed, which leads to the two observed behaviors. In the first behavior, the $R$ cell remains in the low-pressure two-phase region and cannot be correctly updated into the liquid region, which appears as a residual initial discontinuity. In the second behavior, even if the pressure is raised into the liquid region, the increase still lags behind the correct rarefaction structure and propagates to influence downstream cells, manifesting as a non-monotonic density profile.

Below, we further explain this mechanism in detail through the pressure update equation of the RFQC method.

\subsection{RFQC pressure update under fully upwind conditions}

Consider a time $t$ after the initial few steps, at which the downstream cell $R$ of the discontinuity enters the two-phase region due to a pressure rise, and its state is given by:
\begin{equation}
\rho,\quad u,\quad p,\quad q,\quad \xi,\quad E_0,
\end{equation}
with
\begin{equation}
q=\rho e=\xi p+E_0.
\label{eq:qR}
\end{equation}

The state of the upstream liquid cell $L$ is denoted by:
\begin{equation}
\rho_L,\quad u_L,\quad p_L,\quad q_L,\quad \xi_L,\quad E_{0,L},
\end{equation}
subject to:
\begin{equation}
q_L=\rho_L e_L,\quad q_L=\xi_Lp_L+E_{0,L}.
\end{equation}

As cell $R$ enters the two-phase region, the speed of sound drops abruptly to the order of $O(10)$. For the case considered in Fig.~\ref{fig4}, the speed of sound in the liquid phase is $c_L = 130\mathrm{m/s}$; thus, the cell $L$ enters a fully upwind state under the condition of $u_L = 150\mathrm{m/s}$. Without loss of generality, we assume that both cell $L$ and cell $R$ are under a fully upwind condition, which implies that both the left and right interfaces degenerate into one-sided upwind states:
\begin{equation}
\widehat F_{L+1/2}=F_{L},\qquad \widehat F_{R-1/2}=F_{R}.
\end{equation}

Consequently, the updated state of cell $R$ is solely determined by the upstream liquid cell $L$ and itself, independent of any information from cells further downstream.
It is worth noting that under the fully upwind condition, the state of cell $R$ is variable, whereas the state of the liquid cell $L$ is constant. This is because the region upstream of cell $L$ consists entirely of the same liquid phase, while the downstream vapor cell theoretically needs to increase its pressure and density to match the state of cell $L$. Therefore, we explicitly use the subscript $L$ to emphasize its constant state.

\subsubsection{Mass conservation update}

Define the pressure and velocity differences between the two cells:
\begin{equation}
\Delta p=p_L-p,\qquad \Delta u=u-u_L.
\end{equation}

We define the time-step-to-grid-spacing ratio $\theta$ and the update weights $a$ and $b$ as
\begin{equation}
\theta=\frac{\Delta t}{\Delta x},\qquad a=\theta u,\qquad b=\theta u_L.
\end{equation}

Let the time $t+\Delta t$ be denoted by $t^+$, with variables at this time indicated by a superscript "$+$". Under a fully right-going upwind condition, the mass update is given by
\begin{equation}
\rho^+=(1-a)\rho+b\rho_L.
\end{equation}

We introduce two mass weights defined as
\begin{equation}
w_R=(1-a)\rho,\qquad w_L=b\rho_L,\qquad \rho^+=w_R+w_L.
\label{eq:w}
\end{equation}

\subsubsection{$\boldsymbol{\Phi}$ update and pressure update equation}

For a standard RFQC discrete update, $\boldsymbol{\Phi}=(\xi,E_0)$ is updated by upwind mixing:
\begin{equation}
\widetilde \xi^+=(1-b)\xi+b\xi_L,
\label{eq:xi_temp}
\end{equation}
\begin{equation}
\widetilde E_0^+=(1-b)E_0+b E_{0,L}.
\label{eq:E0_temp}
\end{equation}
Here, the tilde symbol $\widetilde{\cdot}$ denotes the values of the thermodynamic variables $\xi$ and $E_0$, as well as the conservatively updated internal energy $q$ discussed below, after a single-step update but prior to re-projection.

In terms of the above variables, the \textbf{single-step pressure-update equation} can be written as:
\begin{equation}
p^+-p = \frac{b\xi_L \Delta p -\theta \Delta u  \left[ q+ \frac{w_Lp+ w_Rp_L}{\rho^+} \right]+\frac{w_Lw_R}{2\rho^+}\Delta u^2 -\frac{\theta^2 \Delta p^2}{2\rho^+}}{\widetilde \xi^+}.
\label{eq:dp2}
\end{equation}

This equation represents the pressure update within cell $R$ under fully upwind conditions. Since the pressure within cell $R$ increases monotonically in LUA problems, we refer to this equation as the single-step \textbf{pressure-increment equation}. The detailed derivation of this equation is provided in \ref{app:dp}. The pressure increment can be decomposed into four terms:

\begin{equation}
p^+-p=Y_b+Y_s+Y_K+Y_H,
\end{equation}
where
\begin{equation}
Y_b=\frac{b\xi_L(p_L-p)}{\widetilde \xi^+},
\label{eq:Yb}
\end{equation}
is the thermodynamic flux-induced pressure-increment term (flux increment term), which represents the capability of the upstream liquid thermodynamic stencil $\boldsymbol{\Phi}$ to lift the pressure of the current cell toward the upstream pressure.

\begin{equation}
Y_s= -\frac{\theta \Delta u}{\widetilde \xi^+} \left[ q+ \frac{w_Lp+ w_Rp_L}{\rho^+} \right],
\end{equation}
is the velocity-difference pressure-work term, which quantifies the negative contribution to the pressure increment associated with the conversion of the pressure difference into a velocity difference.

\begin{equation}
Y_K= \frac{1}{\widetilde \xi^+}\frac{w_Lw_R}{2\rho^+}\Delta u^2,
\end{equation}
is the kinetic-energy mixing term, which describes the pressure increment resulting from the kinetic-energy difference produced when two states with different velocities are mixed.

\begin{equation}
Y_H= -\frac{1}{\widetilde \xi^+}\frac{\theta^2(p_L-p)^2}{2\rho^+},
\end{equation}
is the quadratic pressure-impulse term. It represents the negative contribution to the pressure increment arising from the conversion of the pressure difference into kinetic energy.

\begin{figure}[htbp]
\centering
\includegraphics[width=0.5\textwidth]{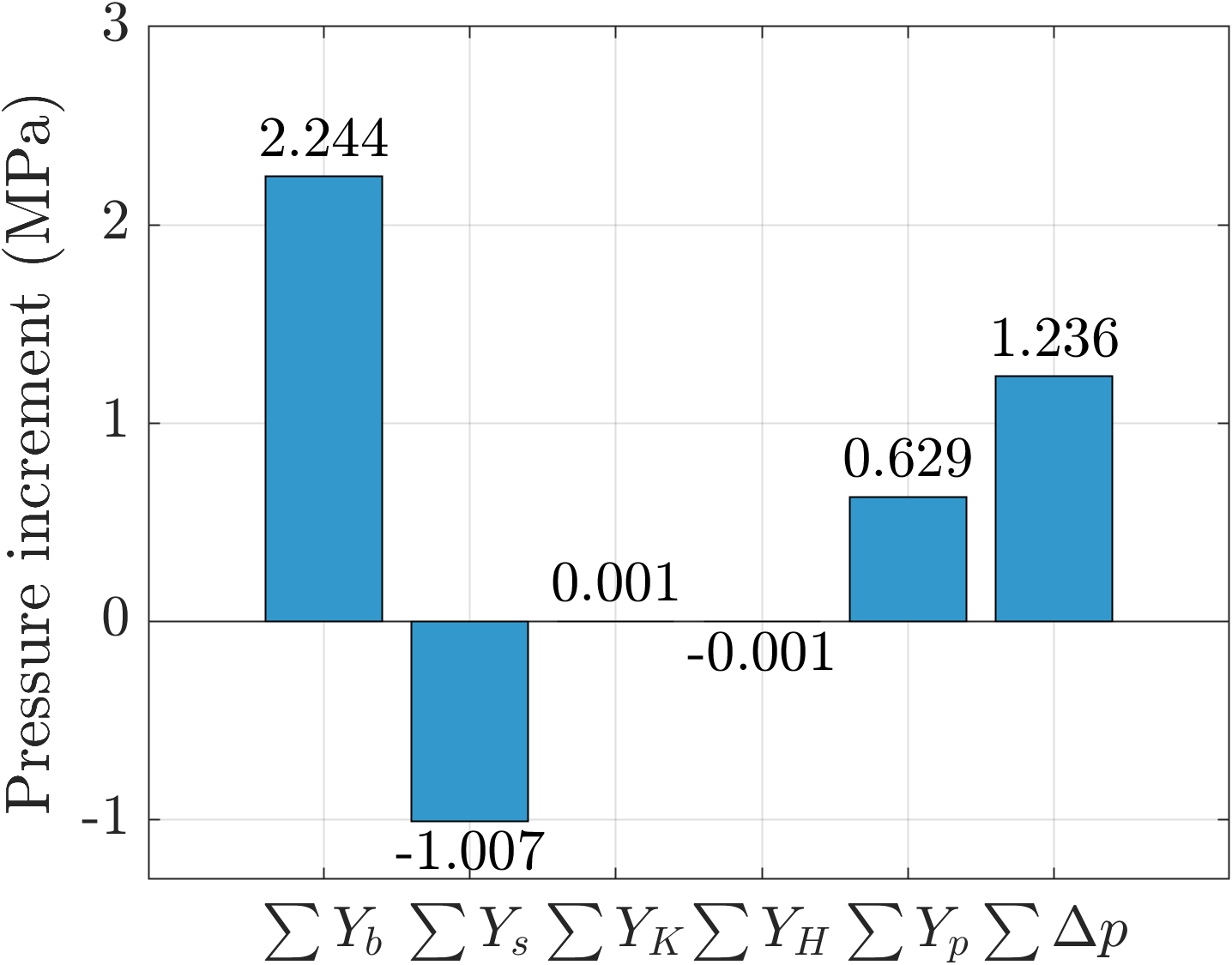}
\caption{Contributions to the pressure increment in cell $R$ for the liquid-upwind anomaly case over $t = 0.6\mathrm{ms}$. The initial translational velocity is $u =150\mathrm{m/s}$, the grid size is $N=500$, and $\text{CFL} =0.5$.}
\label{fig8}
\end{figure}

Figure~\ref{fig8} presents the contributions to the pressure increment in cell $R$ for the LUA case with $u=150\mathrm{m/s}$ shown in Fig.~\ref{fig4}. The numerical results indicate that $Y_b$ is the dominant positive contribution in the anomalous cell $R$, whereas $Y_s$ is the dominant negative contribution. The kinetic-energy mixing term $Y_K$ and the quadratic pressure-impulse term $Y_H$ are smaller by orders of magnitude. Thus, the single-step pressure increment in cell $R$ is governed primarily by the flux increment term $Y_b$. The total pressure increment is only $1.236\mathrm{MPa}$, substantially below the $1.9\mathrm{MPa}$ required to establish a continuous rarefaction wave. This result is consistent with the incomplete expansion of the rarefaction wave observed in Figs.~\ref{fig4}, \ref{fig5}, and \ref{fig6}.

The term $Y_p$ in the figure denotes the positive pressure increment removed by thermodynamic re-projection and will be analyzed in the next subsection.

\subsubsection{Direct source of insufficient pressure rise}

Under fully upwind conditions, the upstream cell $L$ remains in the constant liquid state:
\begin{equation}
p_L = \text{const},\xi_L = \text{const}.
\end{equation}

\begin{figure}[htbp]
\centering
\includegraphics[width=0.98\textwidth]{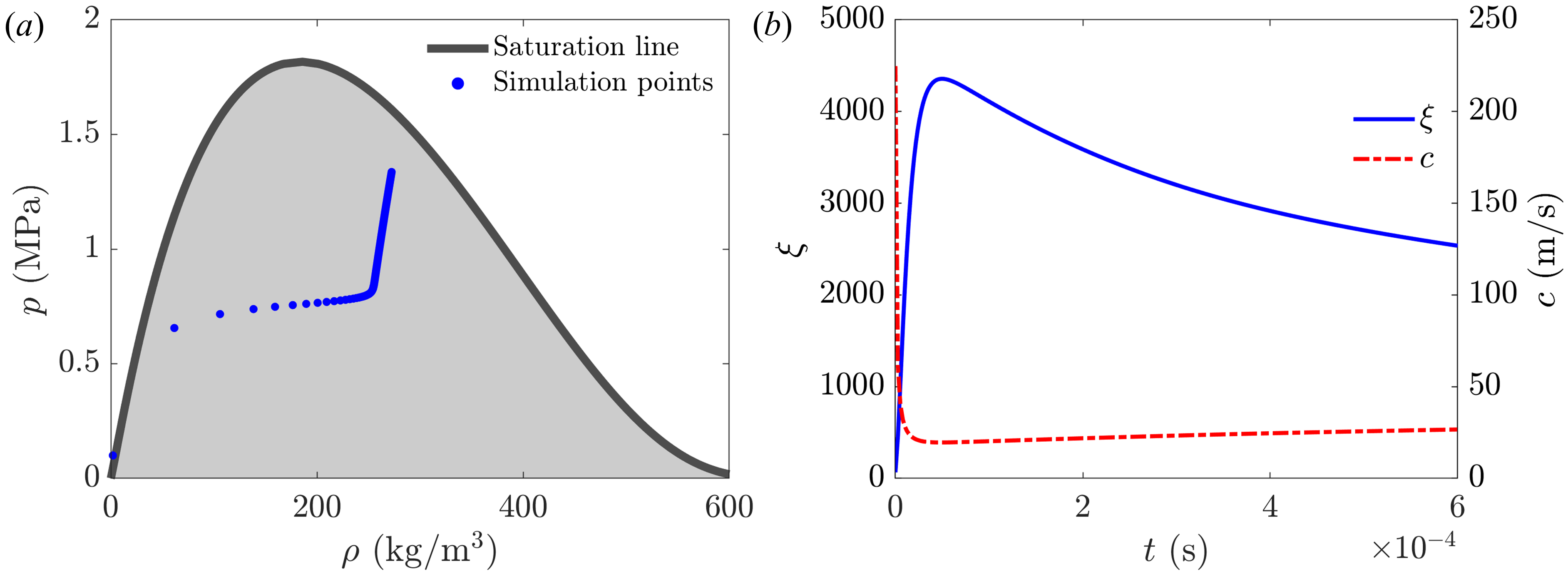}
\caption{(a) State points of the $R$ cell on the $p-\rho$ phase diagram; (b) Temporal variations of $\xi$ and $c$ in the $R$ cell. Parameter variations in the $R$ cell over $t = 0.6\mathrm{ms}$ for the liquid-upwind anomaly case. The initial translational velocity $u =150\mathrm{m/s}$, grid size $N=500$ and $\text{CFL} =0.5$.}
\label{fig9}
\end{figure}

After the initial few time steps (at time $t$), cell $R$ enters the two-phase region, where $\xi$ reaches the order of $O(10^3)$. By contrast, the upstream liquid generally has $\xi=O(10)$. Hence,
\begin{equation}
\xi\gg \xi_L.
\end{equation}
At the same time, the CFL restriction $\Delta t< \theta \Delta x /\text{max}(c+u)$ generally gives $b=\theta u_L< 0.5$. It follows that
\begin{equation}
\widetilde \xi^+=(1-b)\xi+b\xi_L \gg \xi_L
\end{equation}
is primarily governed by the value of $\xi$ in cell $R$. The dominant pressure-increment term therefore reduces to
\begin{equation}
Y_b = \frac{b\xi_L(p_L-p)}{(1-b)\xi+b\xi_L}\simeq \frac{b\xi_L}{(1-b)\xi} (p_L-p).
\end{equation}
Since the coefficient $\frac{b\xi_L}{(1-b)\xi}$ is very small, the pressure increment actually introduced into cell $R$ remains small even when $(p_L-p)$ is large. Furthermore, as $p$ in cell $R$ rises over successive time steps, $(p_L-p)$ decreases accordingly. Since $\xi$ changes only slightly within the two-phase region, $Y_b$ decreases continuously as $p$ rises. Thus, as the pressure in cell $R$ approaches the upstream pressure, the pressure increment produced in each subsequent time step becomes progressively smaller. As a result, cell $R$ remains trapped in the two-phase region for a long period. Figure~\ref{fig9} shows that the solution points of cell $R$ become concentrated in the phase diagram, indicating that the pressure-rise rate decreases progressively. Meanwhile, cell $R$ persists in the two-phase region, where $\xi$ stays at the order of $O(10^3)$ and varies only slightly.

\subsection{Thermodynamic re-projection error and the lost pressure increment}

Thermodynamic re-projection constitutes a central step of the RFQC method. Following pressure recovery, RFQC re-projects the thermodynamic state at the end of each time step. The updated density and pressure, $\rho^+$ and $p^+$, are used to recalculate $q^+$, $\xi^+$, and $E_0^+$, ensuring that $\boldsymbol{\Phi}$ evolves consistently and that the Abgrall condition is satisfied. The internal-energy error introduced by re-projection is defined as:
\begin{equation}
\epsilon_p =q_{\rm EOS}(\rho^+,p^+) - \widetilde q^+ = q^+- \widetilde q^+ .
\end{equation}

\begin{figure}[htbp]
\centering
\includegraphics[width=0.95\textwidth]{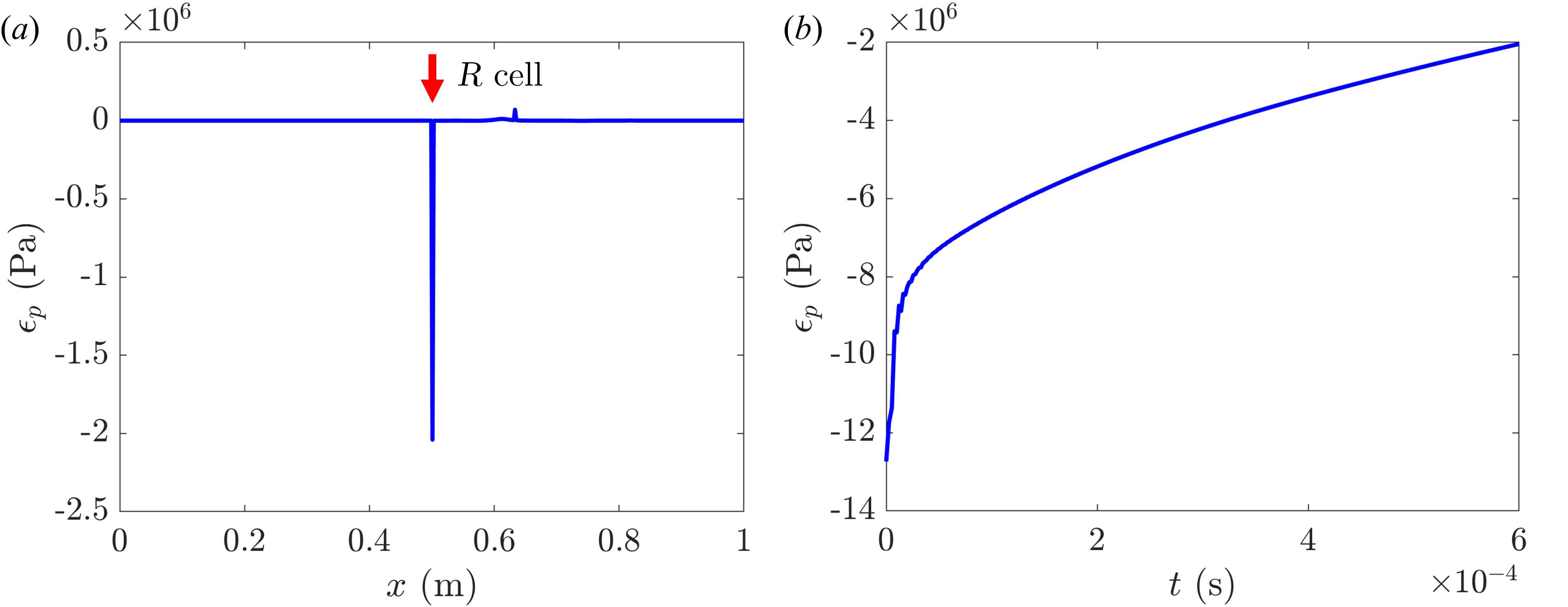}
\caption{(a) Distribution of the re-projection error $\epsilon_p$ at $t = 0.6\mathrm{ms}$; (b) Temporal variations of re-projection error $\epsilon_p$ in the $R$ cell. Behavior of the re-projection error in the liquid-upwind anomaly case. The initial translational velocity $u =150\mathrm{m/s}$, grid size $N=500$ and $\text{CFL} =0.5$.}
\label{fig10}
\end{figure}

As revealed by our previous study, within a correctly developed continuous rarefaction wave, $\epsilon_p$ decreases as the wave propagates and spreads, and remains a small error with second-order convergence with respect to the time-step (Section 3 of Ref.~\cite{Bai2026RFQC}). However, this property holds only for smooth cells that evolve normally. In the anomalous cell $R$, where the LUA leaves a residual discontinuity, the re-projection error remains at $O(10^6)$ throughout the evolution, as shown in Fig.~\ref{fig10}. Moreover, in cell $R$, $\epsilon_p$ satisfies
\begin{equation}
\epsilon_p\ll 0.
\end{equation}

Now, we will demonstrate that, without thermodynamic re-projection, this internal energy error would contribute a positive pressure increment in the next pressure-recovery step. By removing the excess energy residual, thermodynamic re-projection equivalently eliminates a part of the pressure increment required to correctly raise the pressure.

\subsubsection{Pressure increment removed by thermodynamic re-projection}

Note that Eq.~\ref{eq:dp2} is not valid for an arbitrary internal energy $q$; rather, it is derived specifically for a thermodynamically closed state after re-projection. If the internal energy is not re-projected at the end of the time step, the state becomes:
\begin{equation}
q^{c}=\xi p+E_0-\epsilon_p=q-\epsilon_p.
\end{equation}
We emphasize again that $q^c$ cannot be substituted directly into Eq.~\ref{eq:dp2}, because the closure relation in Eq.~\ref{eq:qR} (with the implicit assumption $\epsilon_p=0$) was used in its derivation.

Since cell $L$ remains constant under fully upwind conditions, its re-projection error satisfies $\epsilon_{p,L}=0$. By contrast, the re-projection error $\epsilon_p$ in cell $R$ retained from the preceding time step enters the next update with the geometric weight $(1-a)$. The temporary internal-energy update in the next time step therefore becomes

\begin{equation}
\widetilde q^{+,c} = \widetilde q^+ - (1-a)\epsilon_p.
\end{equation}
Note that $\widetilde E_0^+$ and $\widetilde\xi^+$ are still obtained from the upwind advection update of $\boldsymbol{\Phi}$ (Eqs.~\ref{eq:xi_temp} and \ref{eq:E0_temp}.) The pressure recovery then becomes
\begin{equation}
p^{+,c} = \frac{\widetilde q^+ - (1-a)\epsilon_p - \widetilde E_0^+}{\widetilde\xi^+}.
\end{equation}
It follows that
\begin{equation}
p^{+,c} = p^+ - \frac{(1-a)\epsilon_p}{\widetilde\xi^+}.
\end{equation}
The extra term on the right-hand side is the pressure increment removed by thermodynamic re-projection. The CFL restriction ensures that $1-a>0$, and since $\epsilon_p <0$, the corresponding pressure-increment term is therefore positive:
\begin{equation}
Y_p=  -\frac{(1-a)\epsilon_p}{\widetilde\xi^+} >0.
\end{equation}

Thus, the re-projection step in the RFQC method removes the positive pressure-increment term $Y_p$, resulting in an insufficient total pressure increment in cell $R$. As shown in Fig.~\ref{fig8}, the numerical results indicate that the re-projection pressure-increment term $Y_p$ is comparable in magnitude to the flux increment term $Y_b$. If the pressure increment removed by re-projection is added back to the total pressure increment, then $\sum \Delta p+\sum Y_p = 1.865 \mathrm{MPa}$, which is very close to the $1.9 \mathrm{MPa}$ required to form a continuous rarefaction wave.

Since the anomalous cell $R$ remains trapped in the low-pressure two-phase region, the pressure difference between the two cells $\Delta p=p_L-p$ remains substantial. Consequently, the re-projection error in this discontinuous cell does not exhibit the second-order convergence characteristic typical of smooth regions. Instead, as time steps advance, the error satisfies $-\epsilon_p \gg 0$ (Fig.~\ref{fig10}), which is regenerated by the re-projection at the next time step, equivalently removing the positive pressure increment $Y_p$ once again. A closed feedback loop is thus established:

\begin{align*}
    p  ~\text{rises sluggishly}& \rightarrow R\text{ remains in the two-phase region} \rightarrow Y_p \propto -\epsilon_p \gg 0 \\
    &\rightarrow \text{re-projection removes } Y_p \rightarrow p^+ ~\text{rises sluggishly}.
\end{align*}

In conclusion, while the thermodynamic reprojection constitutes an indispensable closure for RFQC to preserve pressure equilibrium, it repeatedly removes the positive pressure increment in the anomalous cell, allowing the erroneous state to persist.

\subsection{Explanation of the characteristics of the liquid-upwind anomaly}

Thus far, through the analytical derivation of the pressure increment equation, we have elucidated the mechanism of the LUA. Specifically, in the presence of phase transition,the pressure in cell $R$ rises far more slowly than required by the correct wave structure. On the one hand, once cell $R$ enters the two-phase region, the speed of sound decreases rapidly and $\xi$ increases sharply, so that the pressure increment introduced by the upstream liquid flux becomes very limited. Under fully upwind and phase-change conditions, the dominant positive term in the pressure increment equation,
\begin{equation*}
Y_b = \frac{b\xi_L(p_L-p)}{(1-b)\xi_R+b\xi_L}\simeq \frac{b\xi_L}{(1-b)\xi_R} (p_L-p),
\end{equation*}
is severely deficient due to the constraint $\xi_R \gg \xi_L$. On the other hand, thermodynamic re-projection is performed at every time step and equivalently removes the local positive increment term:
\begin{equation*}
     Y_p=  -\frac{(1-a)\epsilon_p}{\widetilde\xi^+} >0.
\end{equation*}
The pressure difference $\Delta p$ between cells $L$ and $R$ therefore persists. This causes a large re-projection error to be repeatedly generated at the anomalous cell, which continuously eliminates the positive pressure increment and forms a feedback loop. The resulting feedback loop delays the pressure rise in cell $R$ and ultimately produces the LUA.

Based on the mechanism described above, the characteristic behaviors of the LUA presented in Section~\ref{sec:LUA_char} can therefore be consistently accounted for:

\begin{itemize}
\item \textbf{Residual initial discontinuity:} As established above, the pressure in cell $R$ rises too slowly to leave the two-phase region and evolve into the liquid phase. A pressure difference $\Delta p$ therefore persists between cells $R$ and $L$, leaving a residual of the initial discontinuity. This corresponds to the $u=150\mathrm{m/s}$ curve near $x=0.5\mathrm{m}$ in Fig.~\ref{fig4}.
\item \textbf{Non-monotonic density in the rarefaction wave:} As the advection velocity $u_L$ increases, the dominant pressure-increment term $Y_b \propto b=\theta u_L$ also increases, allowing cell $R$ to leave the two-phase region and a continuous rarefaction wave to form. At the same time, however, the entire wave is advected more rapidly, so cell $R$ must complete the required pressure rise within a shorter physical time. The cell may therefore be "unlocked," while its pressure evolution still lags behind the correct rarefaction structure. This mismatch ultimately produces an anomalous non-monotonic density profile, as observed near $x=0.65\mathrm{m}$ for the $u=300\mathrm{m/s}$ curve in Fig.~\ref{fig4}.
\item \textbf{Improved wave profile with an increased CFL number:} Since $Y_b \propto b=\theta u_L$, increasing the CFL number raises $\theta$ and hence $Y_b$. The resulting increase in the single-step pressure increment in cell $R$ brings the wave profile closer to the correct solution (Fig.~\ref{fig6}).
\item \textbf{Improved wave profile under grid refinement:} The term $Y_b$ is proportional to $\theta = \Delta t/ \Delta x$ and does not depend directly on the grid size. However, grid refinement reduces the physical length of each time step. Therefore, the required pressure increment over one step decreases accordingly, the increment produced by RFQC becomes closer to the physically required value, and the wave profile improves.
\item \textbf{Insensitivity to time integration:} Regardless of the Runge-Kutta integration scheme employed, the same extreme condition $\xi_R \gg \xi_L$ as in the first-order temporal update is inevitably encountered within each Runge-Kutta substep, thereby failing to resolve the LUA.
\item \textbf{High-pressure liquid and low-pressure vapor:} The LUA requires a high-pressure phase-changing liquid on one side of the initial discontinuity and a low-pressure vapor on the other. Only then can the vapor cell enter the two-phase region during time stepping, leading to a jump of $\xi$ by orders of magnitude and a sharp drop in the dominant pressure-increment term $Y_b$.
\item \textbf{Global advection with liquid upwind:} In the absence of global advection, or when the liquid is downstream, cell $R$ does not need to be raised to the liquid-phase pressure. It remains in the vapor phase or the two-phase region, while the pressure in cell $L$ and the cells to its left decreases, allowing the continuous rarefaction wave to develop correctly.
\end{itemize}

\section{Remedy of the Liquid-upwind Anomaly}
\label{sec:LUA_reg}

Regularizing the initial conditions currently provides the simplest and most reliable remedy for the liquid-upwind anomaly. Numerical tests indicate that the LUA can be avoided by introducing a gradient over neighboring cells symmetrically on both sides of the initial discontinuity along the advection direction. The regularization prevents the extreme configuration $p_L \gg p_R,\xi_L \ll \xi_R$ from arising at the beginning of the calculation. Instead, the first time step produces $p_{L-1} > p_L > p_R > p_{R+1}$ and $\xi_L \approx \xi_R \gg \xi_{L-1},\xi_{R+1}$. The problematic $L/R$ interface is therefore eliminated. The regularized cells then promote a rapid pressure rise downstream and prevent the loss of pressure increment associated with the re-projection error from persisting, allowing the rarefaction wave to develop correctly.

We propose a $\tanh$-based method for generating regularized initial conditions. Both components of $\mathcal{W}=(\rho,p)$ are smoothed along the spatial grid in the advection direction, producing a gradient over one to three cells on each side of the initial discontinuity. The corresponding thermodynamic affine variables $\boldsymbol{\Phi}(\rho,p)$ are then recalculated. For two initial states, the regularized state is given by
\begin{equation}
\mathcal{W}=\varphi\mathcal{W}_L+(1-\varphi){W}_R
\end{equation}
where the regularization function is
\begin{equation}
\varphi = \frac{1}{2} \left( 1 - \mathrm{tanh}\left( \frac{x - x_0}{\gamma \Delta x} \right) \right)
\end{equation}
Here, $x_0$ is the location of the initial discontinuity, and $\gamma$ is the regularization parameter. Numerical tests for various liquid-upwind phase-change Riemann problems involving n-dodecane suggest $\gamma=0.2\sim 1$. In general, $\gamma$ should be reduced as the initial liquid density increases.

\begin{figure}[h]
\centering
\includegraphics[width=0.95\textwidth]{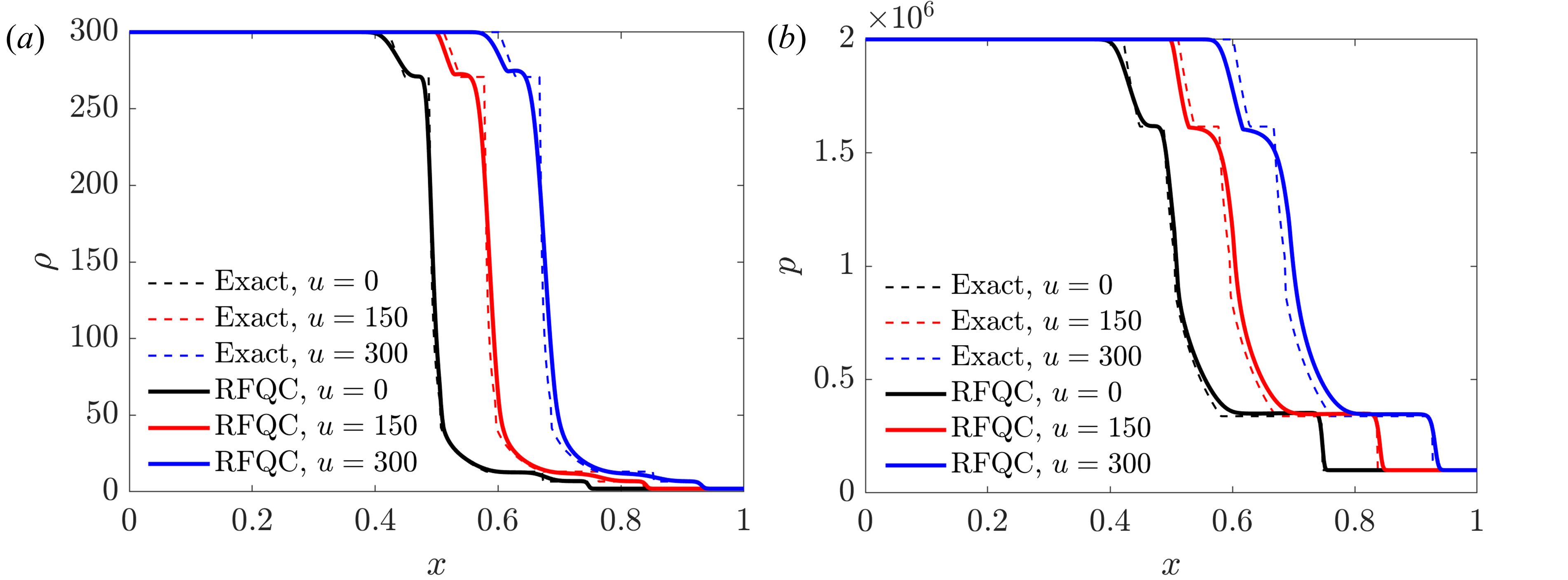} 
\caption{(a) Density; (b) Pressure. RFQC method with smooth initialization. The computational time $t=0.6\mathrm{ms}$, with a grid size of $N=500$ and a CFL number of $0.5$.}
\label{fig11}
\end{figure}

\begin{figure}[h]
\centering
\includegraphics[width=0.95\textwidth]{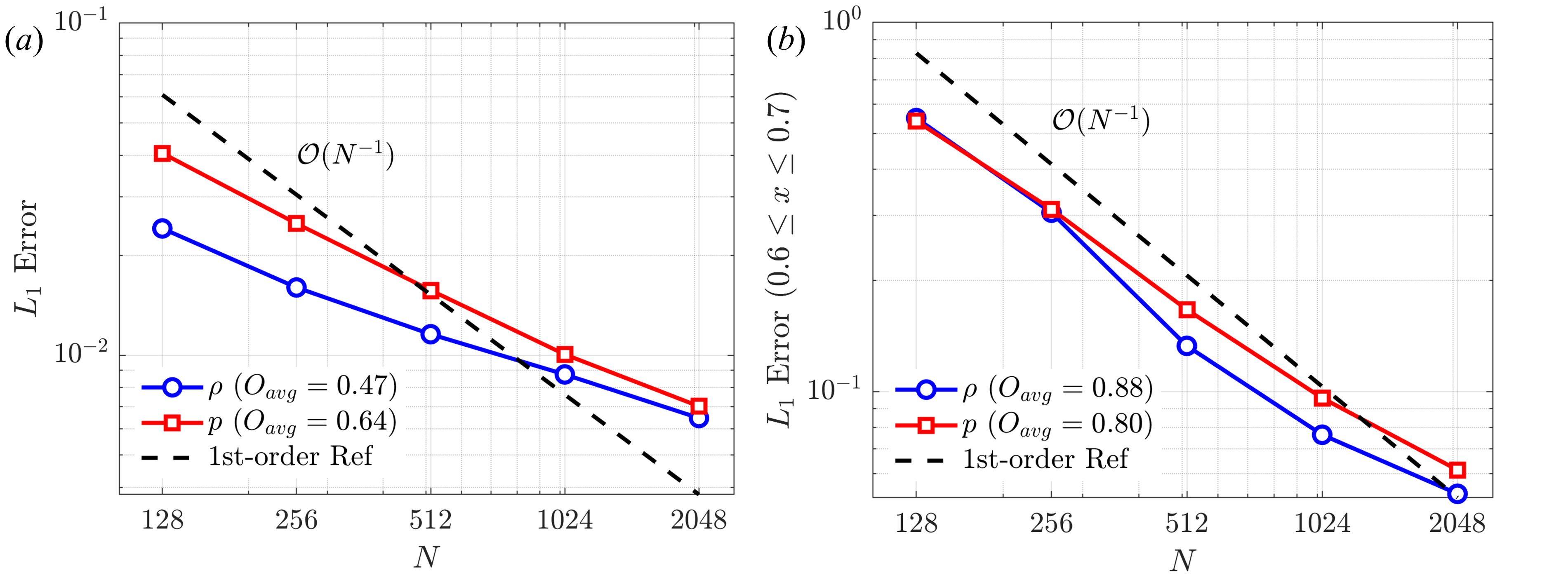} 
\caption{(a) Global $L_1$ error; (b) $L_1$ error in smooth region. Grid-convergence behavior of the RFQC method with smooth initialization. The computational time $t=0.6\mathrm{ms}$, initial translational velocity $u =150\mathrm{m/s}$, $\text{CFL} =0.5$.}
\label{fig12}
\end{figure}

For the cases in Fig.~\ref{fig4} of Section~\ref{sec:LUA_char}, the initial discontinuity is regularized with $\gamma=1$ over $N_s=2$ cells on either side. The resulting profiles are presented in Fig.~\ref{fig11}. The regularization restores the correct wave structure, with the numerical solutions agreeing closely with the exact solutions. We then examine grid convergence for the case with $u=150\mathrm{m/s}$, as shown in Fig.~\ref{fig12}. The regularized solution exhibits smooth convergence curves and stable convergence orders. The overall convergence orders are approximately $0.47$ for density and $0.64$ for pressure, while the corresponding values in the smooth region reach $0.88$ and $0.8$. This behavior is consistent with the expected convergence of a first-order Godunov method.

\begin{figure}[h]
\centering
\includegraphics[width=0.98\textwidth]{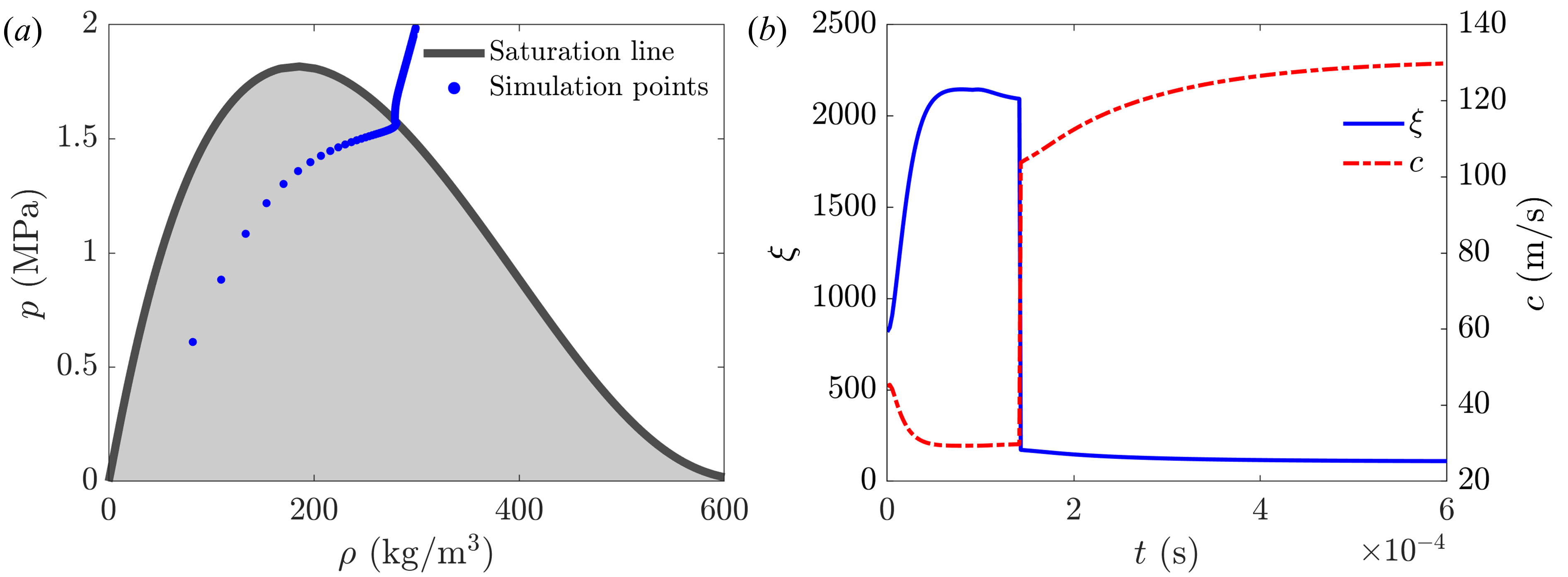}
\includegraphics[width=0.96\textwidth]{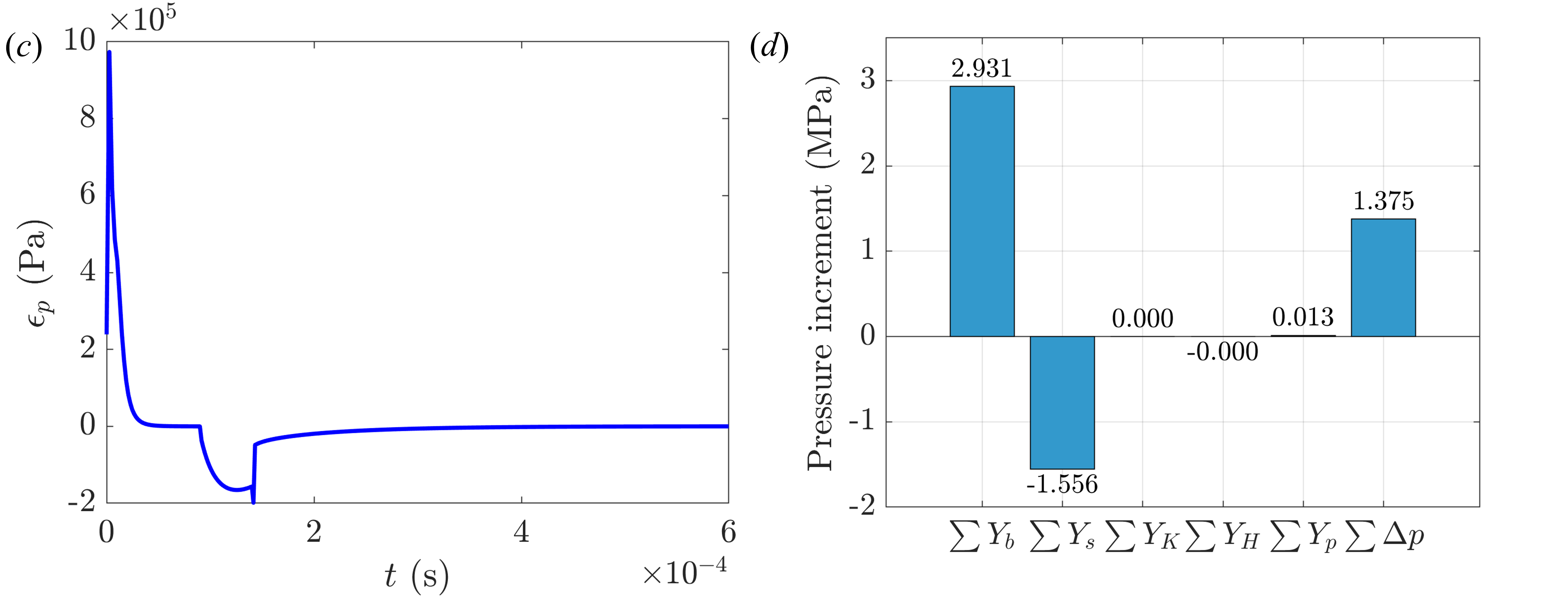}
\caption{(a) State points on the $p-\rho$ phase diagram; (b) Temporal variations of $\xi$ and $c$. (c) Temporal variations of re-projection error $\epsilon_p$ (d) Contributions to the pressure increment. Parameter variations in the $R$ cell over $t = 0.6\mathrm{ms}$ with regularized initialization. The initial translational velocity $u =150\mathrm{m/s}$, grid size $N=500$ and $\text{CFL} =0.5$.}
\label{fig13}
\end{figure}

We further investigate the evolution of the anomalous cell $R$ after regularization, as presented in Fig.~\ref{fig13}. As shown in Fig.~\ref{fig13}(a), the state in cell $R$ evolves smoothly from the two-phase region into the liquid region and reaches the target state. As the state in cell $R$ retrogradely crosses the liquid saturation line, the speed of sound $c$ increases, while $\xi$ decreases sharply(Fig.~\ref{fig13}(b)). Figure~\ref{fig13}(c) shows the temporal evolution of the re-projection error $\epsilon_p$ in cell $R$. The error decreases rapidly and eventually converges to nearly zero. The pressure-increment distribution in cell $R$ (Fig.~\ref{fig13}(d)) further shows that regularization reduces the pressure-increment loss caused by thermodynamic re-projection to only $0.013\mathrm{MPa}$, compared with $0.629\mathrm{MPa}$ without regularization. This further confirms that regularization eliminates the extreme initial condition $p_L \gg p_R,\xi_L \ll \xi_R$, avoids the pressure loss caused by the large re-projection error, and ensures the correct evolution of the wave profile.

Finally, we briefly comment on high-order reconstruction in the presence of the LUA. The current high-order reconstruction follows the method of Johnsen \cite{Johnsen2006} and reconstructs the primitive variables $W=(\rho, u, p, \xi, E_o)$. Strictly speaking, a direct analysis of the pressure update under high-order reconstruction remains difficult. For real fluids, polynomial-based high-order reconstruction can preserve pressure equilibrium,but it is not strictly thermodynamically consistent, because the reconstructed quantities generally do not satisfy $\boldsymbol{\Phi}_{i+1/2}=\boldsymbol{\Phi}(\rho_{i+1/2},p_{i+1/2})$. High-order reconstruction may slightly alleviate the LUA, but the extent of this improvement cannot be predicted quantitatively. Consequently, for practical engineering simulations with initial conditions that may trigger the LUA, initial-condition regularization should be performed first, followed by verification of the corresponding one-dimensional solution before any two- or three-dimensional simulation is carried out. The next section illustrates this workflow with a two-dimensional sonic phase-change jet problem.

\section{Two-dimensional Sonic Phase-change Jet}

The triggering of the LUA typically manifests in the initial conditions of high-speed phase-change jet simulations. Although liquid phase-change jets in practical engines generally do not reach sonic speeds, we consider a sonic jet in this section to highlight the critical challenge posed by the LUA and to demonstrate the robustness of the RFQC method for such problems. The simulation is performed with the RFQC method combined with initial-condition regularization. Time integration is carried out with the third-order SSP-RK scheme \cite{Shu1988}, while spatial discretization adopts Johnsen's primitive-variable reconstruction method \cite{Johnsen2006}. Further details can be found in our previous work~\cite{Bai2026RFQC}.

\begin{figure}[H]
\centering 
\includegraphics[width=0.7\textwidth]{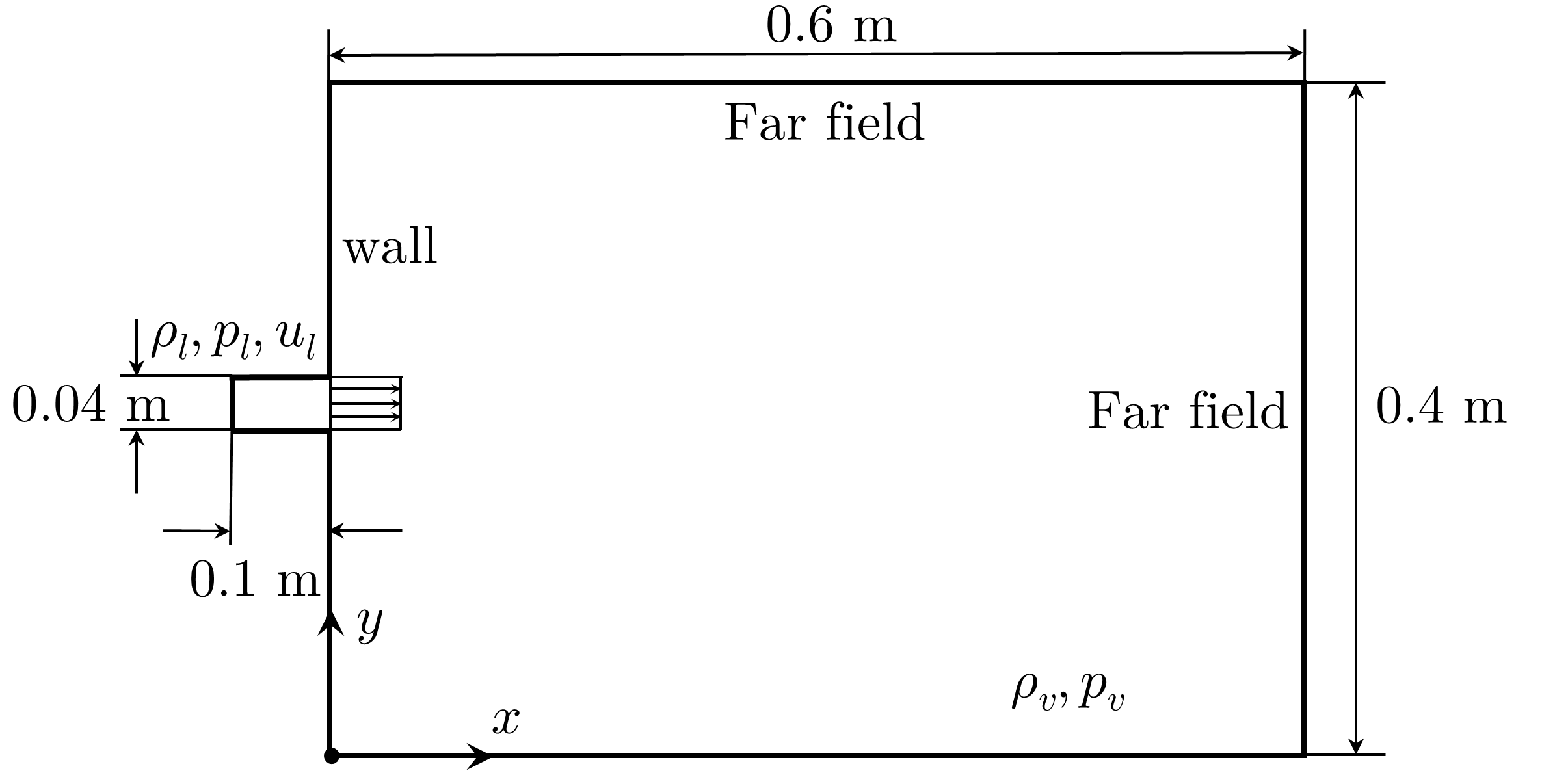} 
\caption{Computational conditions for the sonic phase-change jet.}
\label{fig14}
\end{figure}

Figure~\ref{fig14} shows the computational conditions for the sonic phase-change jet. The computational domain consists of a $0.6\mathrm{m} \times 0.4\mathrm{m}$ vapor-phase region and a $0.1\mathrm{m} \times 0.04\mathrm{m}$ liquid-phase region. Initially, the vapor-phase region has a density of $\rho_{v} = 30 \mathrm{kg/m^3}$, a pressure of $p_v = 1\times 10^5  \mathrm{Pa}$, and zero velocity. The pressure in the liquid-phase region is $p_l = 2\times 10^6  \mathrm{Pa}$. Two different liquid densities are considered to represent fuel injection under different initial-density conditions. For the low-density phase-change jet, the liquid-phase density is $\rho_{l} = 320 \mathrm{kg/m^3}$, and the jet velocity is set to $u_l=150\mathrm{m/s}$. For the high-density phase-change jet, the liquid-phase density is $\rho_{l} = 450 \mathrm{kg/m^3}$, and the jet velocity is set to $u_l=300\mathrm{m/s}$. The computational domain is discretized with a uniform quadrilateral grid of spacing $\Delta = 2 \times10^{-3}\mathrm{m}$, and the simulation time is $t=1 \mathrm{ms}$.

It should be noted that, although a phase-change jet was also simulated in Section 5 of our previous work \cite{Bai2026RFQC}, the LUA did not occur in that case. This is because the initial pressure in the vapor-phase region was set above the critical pressure. Consequently, during the pressure-rise process, the downstream vapor-phase region did not need to undergo compression from the vapor phase through the two-phase region and into the liquid phase. Therefore, the extreme initial condition $p_L \gg p_R,\xi_L \ll \xi_R$ did not arise during the early stage of the calculation. By contrast, in the cases considered in this section, the initial pressure in the downstream vapor-phase region is far below the critical pressure, representing typical LUA conditions.

\subsection{Low-density sonic phase-change jet}

In this case, the liquid-phase region has a density of $\rho_{l} = 320 \mathrm{kg/m^3}$ and a pressure of $p_l = 2\times 10^6  \mathrm{Pa}$. With the corresponding liquid speed of sound $c_l=147\text{m/s}$, the jet velocity is therefore set to $u_l=150\mathrm{m/s}$.

\begin{figure}[htpb]
\centering 
\includegraphics[width=0.95\textwidth]{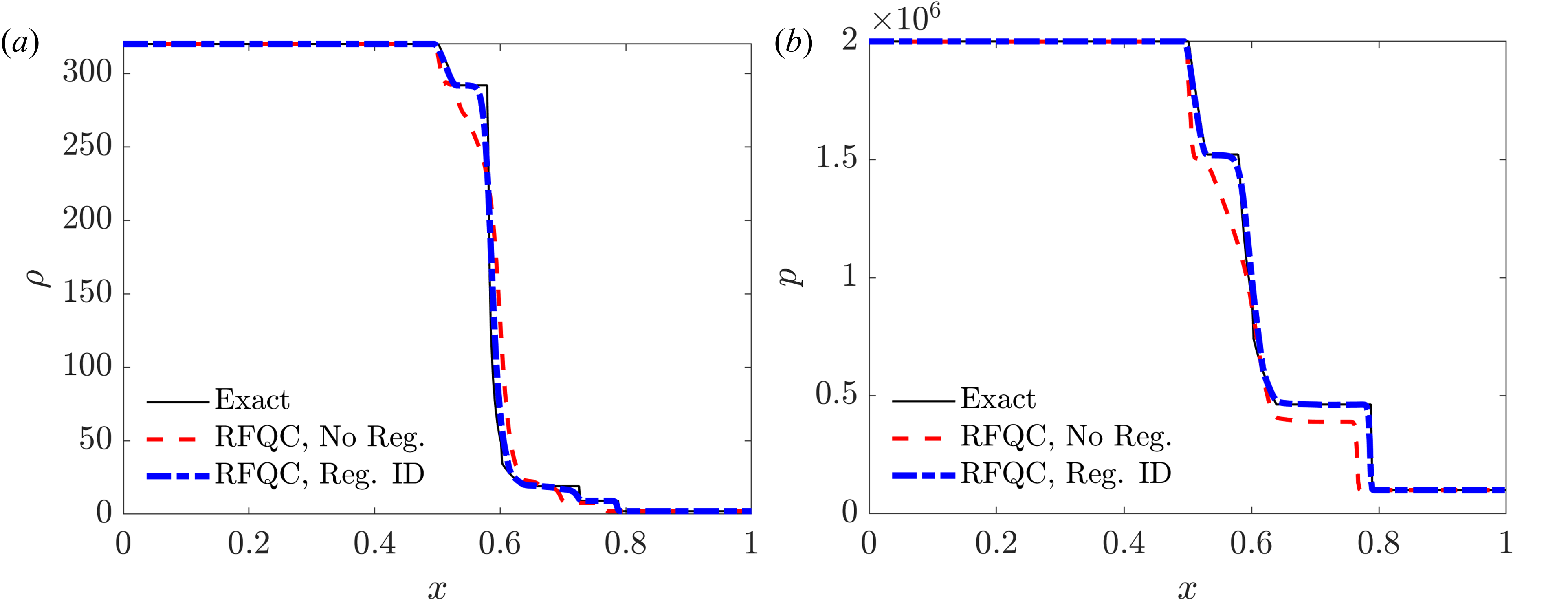} 
\caption{(a) Density,  (b) Pressure. Riemann problem for a low-density phase-change sonic jet: comparison without regularization (No Reg.) and with regularization of the initial discontinuity (Reg. ID). Regularization factor $\gamma = 1.0$, third-order WENO reconstruction is employed for spatial discretization.}
\label{fig15}
\end{figure}

Before performing the two-dimensional jet simulation, we first solve the one-dimensional Riemann problem corresponding to the initial discontinuity in the two-dimensional configuration. The left state of the Riemann problem is set to the liquid state $(\rho_l,u_l,p_l)$, while the right state is set to the vapor state $(\rho_v,u_v,p_v)$. The computational domain has a length of $L=1\mathrm{m}$, the simulation time is $t=0.6\mathrm{ms}$, and the initial discontinuity is located at $x=0.5\mathrm{m}$. A third-order WENO scheme \cite{Jiang1996} is adopted for spatial reconstruction. The CFL number is set to $0.4$, with $N=500$ grid cells ($\Delta = 2 \times10^{-3}\mathrm{m}$), $N_s = 2$ transition cells, and a regularization parameter of $\gamma = 1.0$. The results are shown in Fig.~\ref{fig15}.

\begin{figure}[htpb]
\centering 
\includegraphics[width=0.99\textwidth]{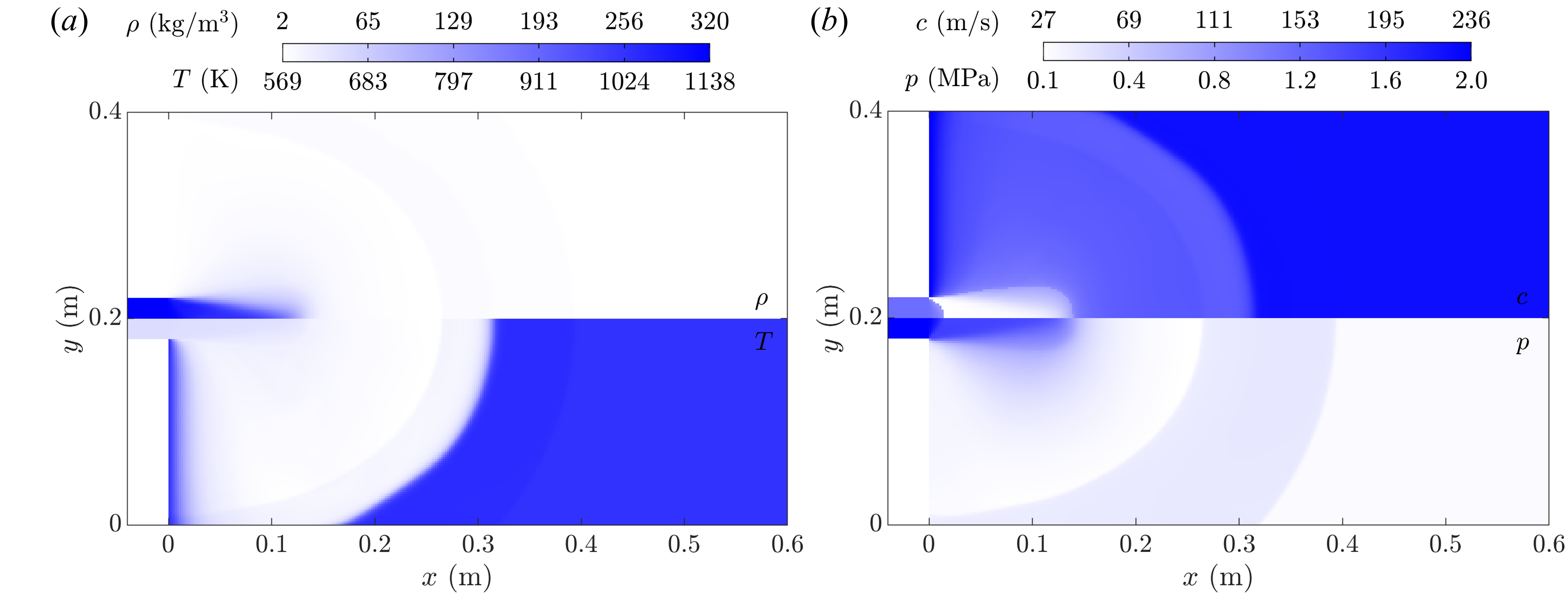} 
\caption{(a) Density and temperature, (b) Sound speed and pressure. Results of the low-density phase-change sonic jet. Regularization factor $\gamma = 1.0$; third-order WENO reconstruction is employed for spatial discretization.}
\label{fig16}
\end{figure}

Without initial-condition regularization, the rarefaction wave fails to evolve correctly, a residual discontinuity remains near $x=0.5\mathrm{m}$, and the downstream pressure is not raised sufficiently. After regularization, the numerical results agree very well with the exact solution. Therefore, the same regularization settings are adopted for the two-dimensional simulation, with $N_s = 2$ transition cells and $\gamma = 1.0$. The CFL number is also retained at $\mathrm{CFL}=0.4$. The simulation results are shown in Fig.~\ref{fig16}.

\subsection{High-density sonic phase-change jet}

In this case, the liquid-phase region has a density of $\rho_{l} = 450 \text{kg/m}^3$ and a pressure of $p_l = 2\times 10^6 \text{Pa}$. Given the corresponding liquid speed of sound $c_l=295\text{m/s}$, the jet velocity is set to $u_l=300\text{m/s}$.

\begin{figure}[htpb]
\centering 
\includegraphics[width=0.95\textwidth]{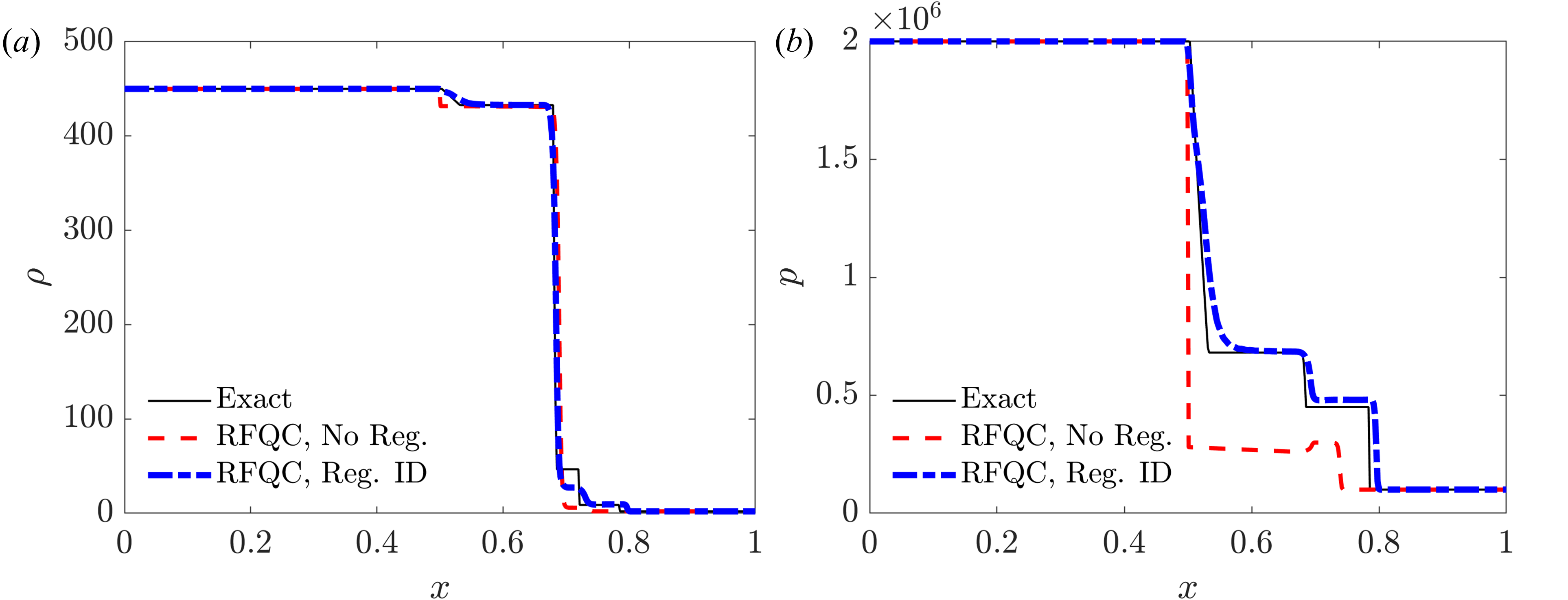} 
\caption{(a) Density,  (b) Pressure. Riemann problem for a high-density phase-change sonic jet: comparison without regularization (No Reg.) and with regularization of the initial discontinuity (Reg. ID). Regularization factor $\gamma = 0.8$, second-order MUSCL reconstruction is employed for spatial discretization.}
\label{fig17}
\end{figure}

Before performing the two-dimensional jet simulation, we first solve the one-dimensional Riemann problem corresponding to the initial discontinuity in the two-dimensional configuration. Similarly, the left and right states are set to $(\rho_l,u_l,p_l)$ and $(\rho_v,u_v,p_v)$, respectively. The computational domain has a length of $L=1\text{m}$, the simulation time is $t=0.6\mathrm{ms}$, and the initial discontinuity is located at $x=0.5\text{m}$. A second-order MUSCL reconstruction with the minmod limiter \cite{Sweby1984} is adopted. The CFL number is set to $0.4$, with $N=500$ grid cells, $N_s = 2$ transition cells, and a regularization parameter $\gamma = 0.8$. The results are shown in Fig.~\ref{fig17}.

\begin{figure}[htpb]
\centering 
\includegraphics[width=0.99\textwidth]{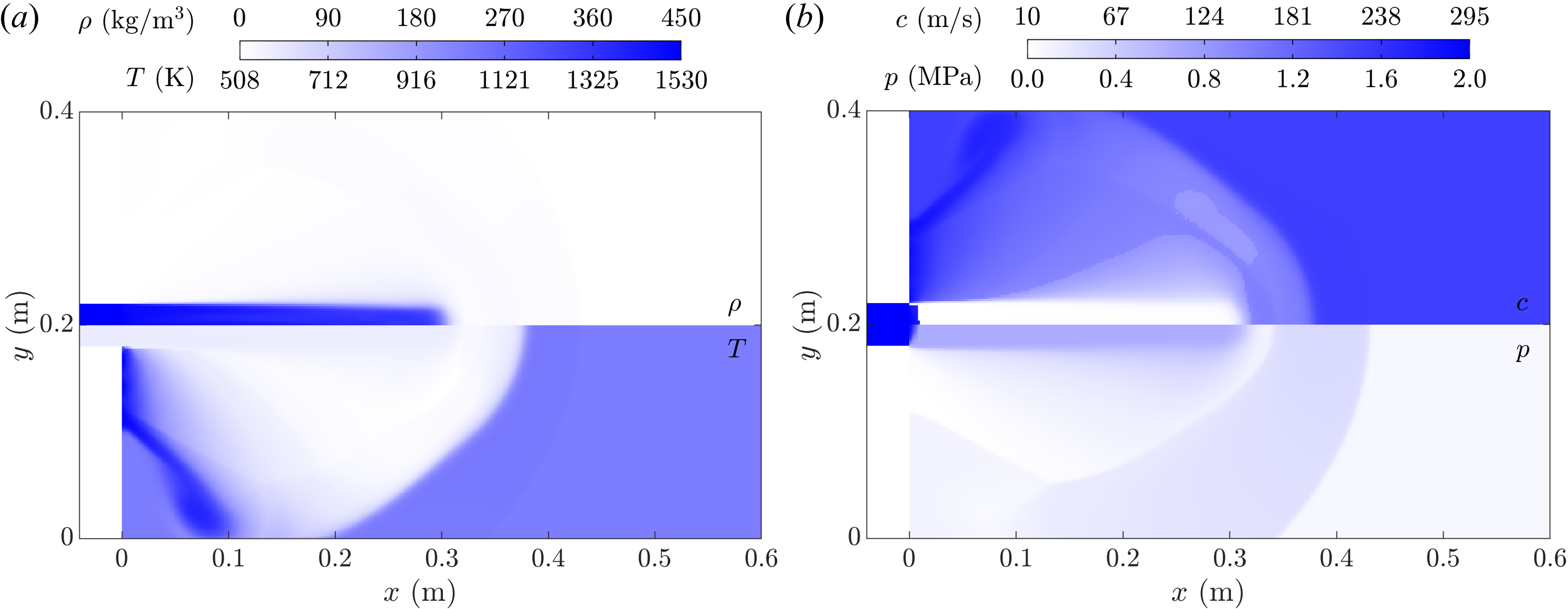} 
\caption{(a) Density and temperature, (b) Sound speed and pressure. Results of the high-density phase-change sonic jet. Regularization factor $\gamma = 0.8$; second-order MUSCL reconstruction is employed for spatial discretization.}
\label{fig18}
\end{figure}

Figure~\ref{fig17} shows that, without initial-condition regularization, the rarefaction wave fails to evolve correctly, a significant residual discontinuity develops near the initial discontinuity, and the downstream pressure remains far below its correct level. After regularization, the numerical results agree well with the exact solution. Therefore, the same regularization settings are adopted for the two-dimensional simulation, with $N_s = 2$ transition cells and $\gamma = 0.8$. The CFL number is also retained at $\text{CFL}=0.4$. The results are shown in Fig.~\ref{fig18}.

\begin{figure}[H]
\centering 
\includegraphics[width=0.95\textwidth]{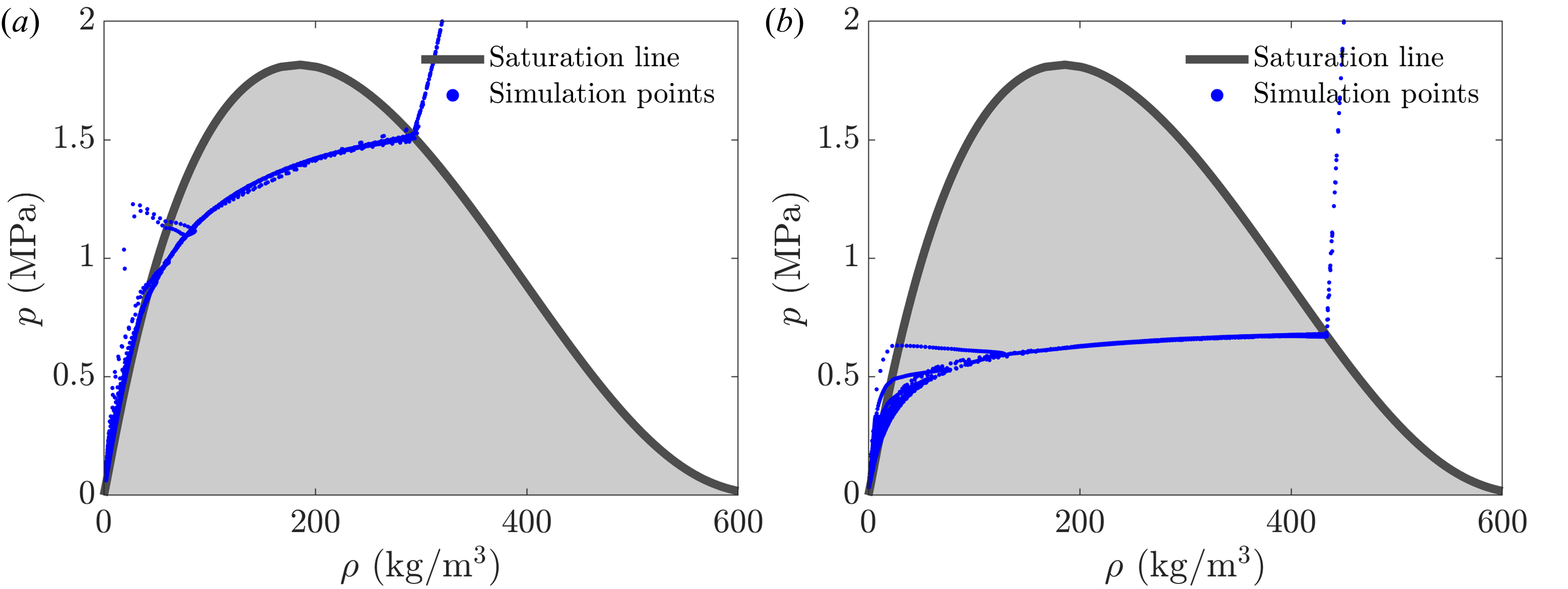} 
\caption{(a) Low-density sonic jet, (b) High-density sonic jet. Simulation points on the $p-\rho$ phase diagram.}
\label{fig19}
\end{figure}

The results demonstrate that, with initial-condition regularization, the RFQC method performs well for sonic phase-change jets. The density, temperature, and pressure within the liquid jet vary continuously, and the discontinuity in the speed of sound is also well captured. The state points from the two jet simulations are further plotted on the phase diagram, as shown in Fig.~\ref{fig19}. In both phase-change jet cases, the isentropes exhibit a kink at the saturation line, indicating a discontinuous change in the speed of sound. For the high-density jet, the solution points within the two-phase region lie almost along a horizontal line, indicating that the speed of sound decreases to the order of $O(1)$.

\subsection{Comparison with the double-flux method}

As discussed above, when solving such complex phase-change flows with high-speed advection, the RFQC method requires the additional step of initial-condition regularization to obtain correct numerical results. However, this does not indicate degraded numerical performance of the RFQC method, but rather reflects the inherent computational difficulties of real multiphase flow problems with phase change. Physically, the fluid interface itself possesses a small but finite thickness due to viscosity and heat transfer. Initial-condition regularization precisely recovers this physical thickness to avoid numerical singularities. It should also be emphasized that the LUA is primarily a start-up anomaly, rather than a phenomenon that would emerge during practical flow evolution. Because physical flow evolutions are governed by dissipation and always tend toward equilibrium, the flow field will not spontaneously generate a discontinuous interface possessing such massive pressure and thermodynamic differentials.

To highlight the challenge of the sonic phase-change jet case for current numerical methods, in this subsection, we simulate the high-density jet considered previously with the double-flux (DF) method, which is currently more widely applied in real-fluid simulations, and compare the results with those obtained by the RFQC method.

Because of stability limitations, both the temporal and spatial discretizations of the DF method are restricted to first order for these cases, and the CFL number is set to $0.2$. All other conditions are identical to those of the RFQC simulations. In principle, the LUA is specific to the RFQC scheme and does not arise in the DF method. Numerical tests also show that initial-condition regularization has almost no influence on the DF results. Therefore, the same regularization is applied to the DF simulations to ensure an identical initial condition.

\begin{figure}[h]
\centering
\includegraphics[width=0.95\textwidth]{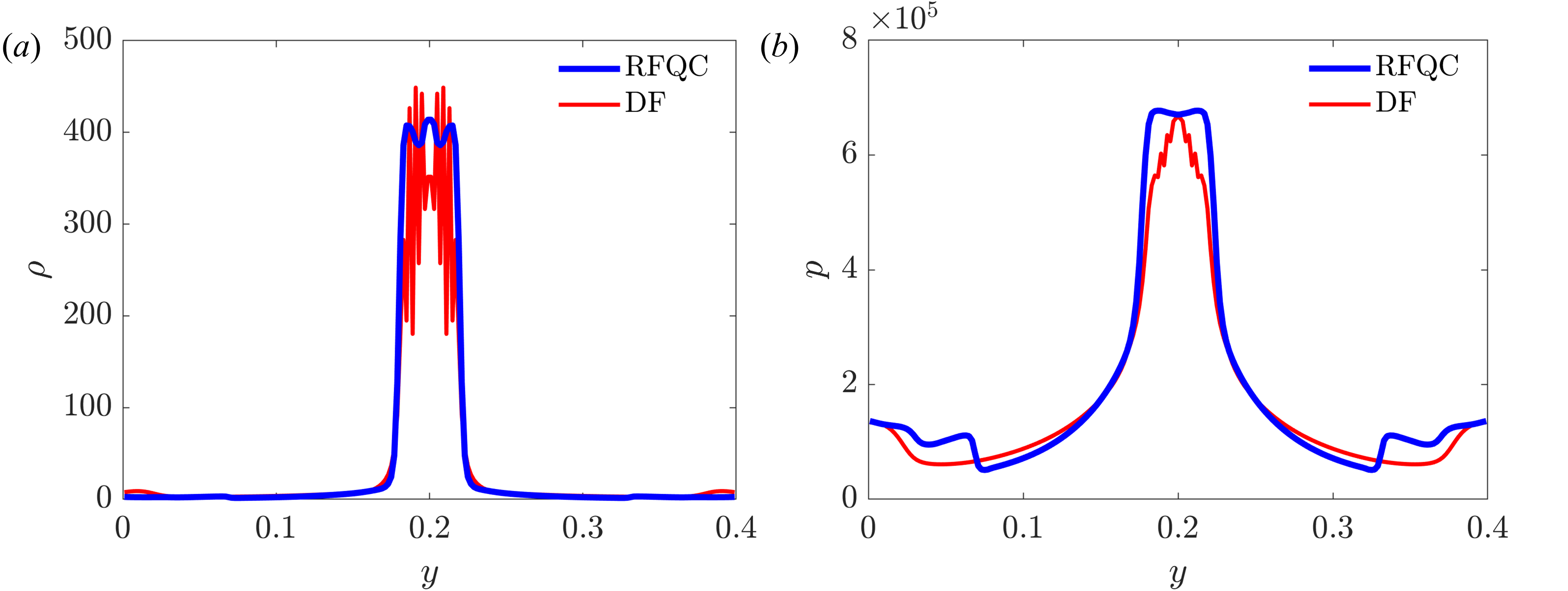}
\includegraphics[width=0.95\textwidth]{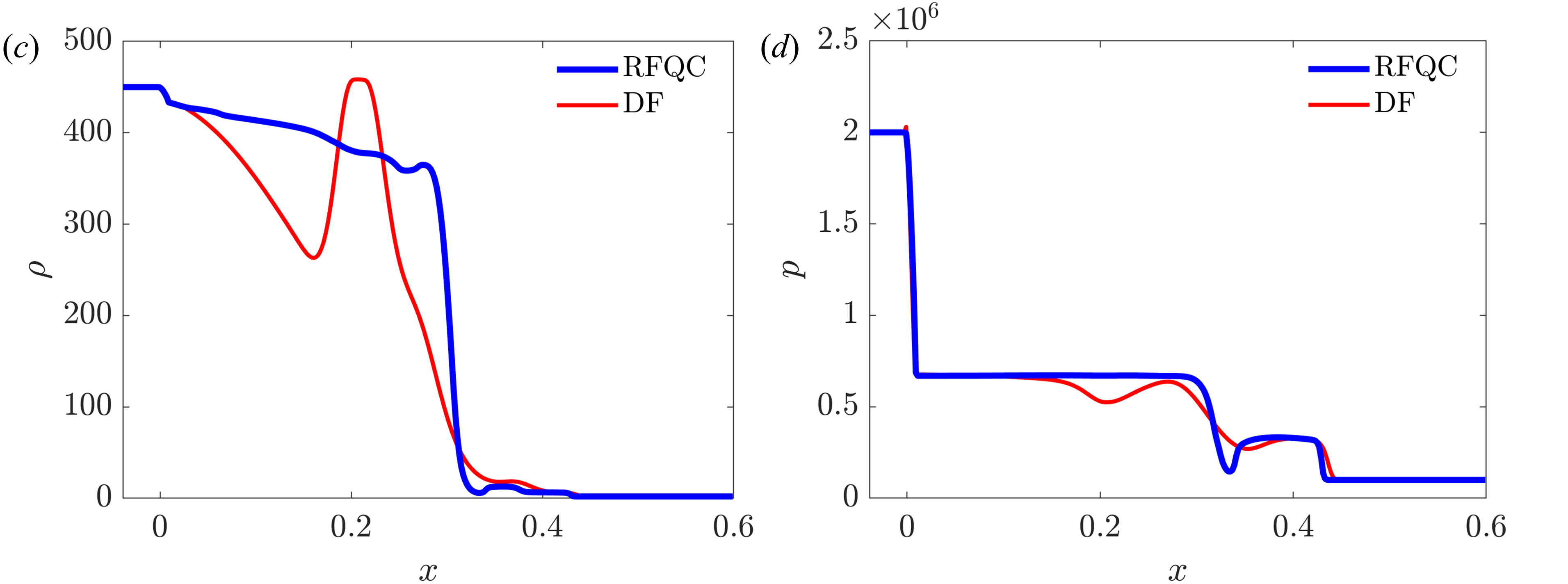}
\caption{(a) Density, $x=0.1\mathrm{m}$; (b) Pressure, $x=0.1\mathrm{m}$; (c) Density, $y=0.2\mathrm{m}$; (d) Pressure, $y=0.2\mathrm{m}$. Comparison of the computational results between the DF and RFQC methods. The DF method utilizes first-order spatio-temporal discretization with a CFL number of 0.2.}
\label{fig20}
\end{figure}

The numerical results are extracted along the section at $x=0.1\mathrm{m}$ in the computational domain shown in Fig.~\ref{fig14} (corresponding to a distance of $0.1\mathrm{m}$ from the jet exit) and along the section at $y=0.2\mathrm{m}$ (corresponding to the jet centerline). These results are compared with those obtained by the RFQC method in the previous subsection, as shown in Fig.~\ref{fig20}.

The results show that the RFQC method remains stable with second-order reconstruction of the primitive variables. The pressure and density profiles vary smoothly. The jet shock near $x=0.45\mathrm{m}$ in Fig.~\ref{fig20}(d), and the intercepting shocks near $x=0.35\mathrm{m}$ in Fig.~\ref{fig20}(d) and near $y=0.07\mathrm{m}$ and $y=0.33\mathrm{m}$ in Fig.~\ref{fig20}(b), are clearly resolved. By contrast, the DF method exhibits severe density and pressure oscillations even with first-order discretization (Figs.~\ref{fig20}(a) and \ref{fig20}(b)). This is also why only first-order discretization and a smaller CFL number are adopted for the DF method; otherwise, numerical stability cannot be ensured. Meanwhile, the DF results exhibit an anomalous increase in density along the streamwise direction (Fig.~\ref{fig20}(c)) and pressure fluctuations (Fig.~\ref{fig20}(d)). These results clearly demonstrate the robustness and accuracy of the RFQC method for phase-change jet simulations.

\section{Conclusion}
This paper first revisits the pressure oscillation problem for real fluids within the finite-volume framework from the perspective of continuum thermodynamics. Physically, in traditional conservative finite-volume methods, pressure oscillations arise from the thermodynamic-equilibrium assumption, which is implicit when recovering pressure from conservative variables. Conversely, the RFQC method recovers mechanical-equilibrium pressure by introducing additional mechanical-equilibrium information. Building upon this, we interpret the central physical counterpart of the RFQC method: the affine parameters $(\xi, E_0)$ of the isentropic internal-energy--pressure relation are evolved along the pathlines to recover mechanical-equilibrium pressure, while thermodynamic re-projection converts the deviation of the evolution from the isentropic trajectory into a controllable internal-energy error. This treatment avoids the evaluation of complex speed-of-sound derivatives and preserves robust shock capturing. These insights suggest a real-fluid algorithm design strategy: introducing additional mechanical-equilibrium information into Godunov methods presents a promising avenue for balancing conservation and non-oscillatory evolution.

Thermodynamic re-projection is essential for ensuring thermodynamic consistency in the RFQC method, but it may also induce a numerical anomaly in a class of extreme phase-change--advection Riemann problems. We identify and rigorously analyze this Liquid-Upwind Anomaly (LUA), and elucidate its formation mechanism. In such special Riemann problems, the jump of affine slope $\xi$ by orders of magnitude during phase change creates a singular condition at the initial discontinuity. Consequently, the re-projection error cannot converge normally, leading to an anomalous cell with delayed pressure rise. Regularizing the initial discontinuity with a transition gradient within neighboring cells introduces a quasi-physical transition layer. The initial divergence of the re-projection error can be effectively suppressed by such regularization, thereby restoring the proper evolution of the method.

The numerical results demonstrate that, with initial-condition regularization, the RFQC method remains robust and accurate for the extreme flow represented by a sonic phase-change jet. Evidently, however, the current regularization strategy represents a technical remedy rather than a fundamental solution, as it relies on the empirical selection of a regularization parameter. Consequently, developing a theoretically more rigorous and effective regularization method will constitute the main direction of our future research.

\appendix
\section{Detailed Derivation of the Single-Step Pressure Update Equation}
\label{app:dp}

Under a fully right-going upwind condition, the momentum update is given by:
\begin{equation}
(\rho u)^+=(1-a)\rho u+b\rho_Lu_L+\theta(p_L-p).
\end{equation}
This simplifies to:
\begin{equation}
(\rho u)^+=w_Ru+w_Lu_L+\theta\Delta p .
\end{equation}
Therefore,
\begin{equation}
u^+=\frac{w_Ru+w_Lu_L+\theta\Delta p}{\rho^+}.
\label{eq:u+}
\end{equation}

The total energy update is given by:
\begin{equation}
(\rho e_t)^+=\rho e_t-\theta\left[u(\rho e_t+p)-u_L(\rho_Le_{t,L}+p_L)\right].
\end{equation}
Substituting the definition of total energy yields:
\begin{align}
(\rho e_t)^+
={}& q+\frac12\rho u^2
-\theta u\left(q+\frac12\rho u^2+p\right)
+\theta u_L\left(q_L+\frac12\rho_Lu_L^2+p_L\right) .
\end{align}
Separating the internal energy, kinetic energy, and pressure work terms gives:
\begin{align}
(\rho e_t)^+
={}& (1-a)q+bq_L
+\frac12(1-a)\rho u^2
+\frac12b\rho_Lu_L^2
+\theta(u_Lp_L-up).
\end{align}
Using $w_R=(1-a)\rho$ and $w_L=b\rho_L$ (Eq.~\ref{eq:w}), this can be written as:
\begin{equation}
(\rho e_t)^+=(1-a)q+bq_L+\frac12w_Ru^2+\frac12w_Lu_L^2+\theta(u_Lp_L-up).
\end{equation}

The temporary volumetric internal energy is obtained by subtracting the kinetic energy at the new time from the total energy:
\begin{equation}
\widetilde q^+=(\rho e_t)^+-\frac12\rho^+(u^+)^2 .
\label{eq:q+1}
\end{equation}
From the velocity update equation \ref{eq:u+}, the kinetic energy at the new time can be expressed as:
\begin{equation}
\frac12\rho^+(u^+)^2
=\frac{1}{2\rho^+}\left(w_Ru+w_Lu_L+\theta\Delta p\right)^2 .
\label{eq:rhou2+}
\end{equation}
Substituting Eq.~\ref{eq:rhou2+} into Eq.~\ref{eq:q+1} and simplifying gives the temporary volumetric internal energy:
\begin{equation}
\widetilde q^+=(1-a)q+bq_L+Q,
\end{equation}
where
\begin{equation}
Q = -\theta \Delta u \frac{w_Lp+ w_Rp_L}{\rho^+} +\frac12\frac{w_Lw_R}{\rho^+}\Delta u^2 -\frac{\theta^2 \Delta p^2}{2\rho^+}.
\label{eq:Q}
\end{equation}
The first term stems from the pressure work under the velocity differential, the second term arises from the kinetic energy of velocity mixing, and the third term originates from the quadratic term of the pressure impulse.

The pressure-recovery equation is:
\begin{equation}
p^+=\frac{\widetilde q^+-\widetilde E_0^+}{\widetilde \xi^+}.
\end{equation}

Substituting the thermodynamic coefficient update equations \ref{eq:xi_temp} and \ref{eq:E0_temp}, as well as the internal energy--pressure affine relation \ref{eq:qR}, we have:
\begin{equation}
\widetilde q^+-\widetilde E_0^+ = (1-a)q+b q_L+Q-(1-b)E_0-b E_{0,L}.
\end{equation}
Expanding this expression gives:
\begin{equation}
(1-a)(\xi p+E_0)+b(\xi_Lp_L+E_{0,L})+Q -(1-b)E_0-b E_{0,L}.
\end{equation}
Here, the $E_{0,L}$ terms cancel out, while the $E_0$ terms remain as:
\begin{equation}
(1-a)E_0-(1-b)E_0=(b-a)E_0.
\end{equation}
Thus, we obtain:
\begin{equation}
\widetilde q^+-\widetilde E_0^+ = (1-a)\xi p+b\xi_Lp_L+(b-a)E_0+Q.
\end{equation}

To express this as a pressure increment relative to $p$, we subtract $p\widetilde \xi^+$ from the above equation and divide by $\widetilde \xi^+$, yielding:
\begin{equation}
p^+-p = \frac{\widetilde q^+-\widetilde E_0^+-p\widetilde \xi^+}{\widetilde \xi^+}.
\end{equation}

The numerator can be expanded as:
\begin{equation}
(1-a)\xi p+b\xi_Lp_L+(b-a)E_0+Q -(1-b)\xi p-b\xi_Lp.
\end{equation}
After rearrangement, we obtain:
\begin{equation}
p^+-p = \frac{b\xi_L(p_L-p)+(b-a)(\xi p+E_0)+Q}{\widetilde \xi^+}.
\label{eq:dp1}
\end{equation}
Substituting:
\begin{equation}
    q=\xi p+E_0, \qquad b-a=-\theta \Delta u,
\end{equation}
gives:
\begin{equation}
p^+-p = \frac{b\xi_L\Delta p-\theta \Delta u q+Q}{\widetilde \xi^+}.
\end{equation}

Finally, substituting the expression for $Q$ from Eq.~\ref{eq:Q} yields the \textbf{single-step pressure update formulation}:
\begin{equation}
p^+-p = \frac{b\xi_L \Delta p -\theta \Delta u  \left[ q+ \frac{w_Lp+ w_Rp_L}{\rho^+} \right]+\frac{w_Lw_R}{2\rho^+}\Delta u^2 -\frac{\theta^2 \Delta p^2}{2\rho^+}}{\widetilde \xi^+}.
\end{equation}

\end{document}